\documentclass[times,twocolumn,final]{elsarticle}

\usepackage{medima}
\usepackage{framed,multirow}

\usepackage{amssymb}
\usepackage{latexsym}

\usepackage{url}
\usepackage{xcolor}

\usepackage{hyperref}

\usepackage{multirow}
\usepackage{float} 

\usepackage{bbm, dsfont}

\usepackage{amsmath,amssymb,amsfonts}
\usepackage{algorithmic}
\usepackage{graphicx}
\usepackage{textcomp}
\usepackage{booktabs}
\usepackage{multirow}
\usepackage{graphicx}
\usepackage{subfigure}
\usepackage{array}
\usepackage{url}

\definecolor{newcolor}{rgb}{.8,.349,.1}

\journal{Medical Image Analysis}

\begin{document}

\verso{Yu \textit{et~al.}}

\begin{frontmatter}

\title{HiFi-Syn: Hierarchical Granularity Discrimination for High-Fidelity Synthesis of MR Images with Structure Preservation}

\author[1,2,3,4]{Ziqi \snm{Yu}}
\author[5]{Botao \snm{Zhao}}
\author[6]{Shengjie \snm{Zhang}}
\author[6]{Xiang \snm{Chen}}
\author[1,2,3,4]{Fuhua \snm{Yan}}
\author[6]{Jianfeng \snm{Feng}}
\author[7]{Tingying \snm{Peng}\corref{cor1}}
\author[1,2,3,4]{Xiao-Yong \snm{Zhang}\corref{cor1}}

\cortext[cor1]{Corresponding author: 
zhangxiaoyong@sjtu.edu.cn; \\tingying.peng@tum.de}

\address[1]{Department of Radiology, Ruijin Hospital, Shanghai Jiao Tong University School of Medicine, Shanghai, China}  
\address[2]{Faculty of Medical Imaging Technology, College of Health Science and Technology, Shanghai Jiao Tong University School of Medicine, Shanghai, China}
\address[3]{Clinical Neuroscience Center, Ruijin Hospital, Shanghai Jiao Tong University School of Medicine, Shanghai, China}  
\address[4]{National Engineering Research Center of Advanced Magnetic Resonance Technologies for Diagnosis and Therapy, Shanghai Jiao Tong University, Shanghai, China}  
\address[5]{Ping An Technology (Shenzhen) Co., Ltd., China}  
\address[6]{Institute of Science and Technology for Brain-Inspired Intelligence, MOE Key Laboratory of Computational Neuroscience and Brain-Inspired Intelligence, Fudan University, Shanghai, China}  
\address[7]{Helmholtz AI, Helmholtz zentrum Muenchen, Munich, Germany}  

\begin{abstract}
Synthesizing medical images while preserving their structural information is crucial in medical research. In such scenarios, the preservation of anatomical content becomes especially important. Although recent advances have been made by incorporating instance-level information to guide translation, these methods overlook the spatial coherence of structural-level representation and the anatomical invariance of content during translation. To address these issues, we introduce hierarchical granularity discrimination, which exploits various levels of semantic information present in medical images. Our strategy utilizes three levels of discrimination granularity: pixel-level discrimination using a Brain Memory Bank, structure-level discrimination on each brain structure with a re-weighting strategy to focus on hard samples, and global-level discrimination to ensure anatomical consistency during translation. The image translation performance of our strategy has been evaluated on three independent datasets (UK Biobank, IXI, and BraTS 2018), and it has outperformed state-of-the-art algorithms. Particularly, our model excels not only in synthesizing normal structures but also in handling abnormal (pathological) structures, such as brain tumors, despite the variations in contrast observed across different imaging modalities due to their pathological characteristics. The diagnostic value of synthesized MR images containing brain tumors has been evaluated by radiologists. This indicates that our model may offer an alternative solution in scenarios where specific MR modalities of patients are unavailable. Extensive experiments further demonstrate the versatility of our method, providing unique insights into medical image translation.

\end{abstract}


\end{frontmatter}


\section{Introduction}
\label{sec:introduction}

\begin{figure}[ht]
  \centering
  \centerline{ \includegraphics[width=0.98\linewidth]{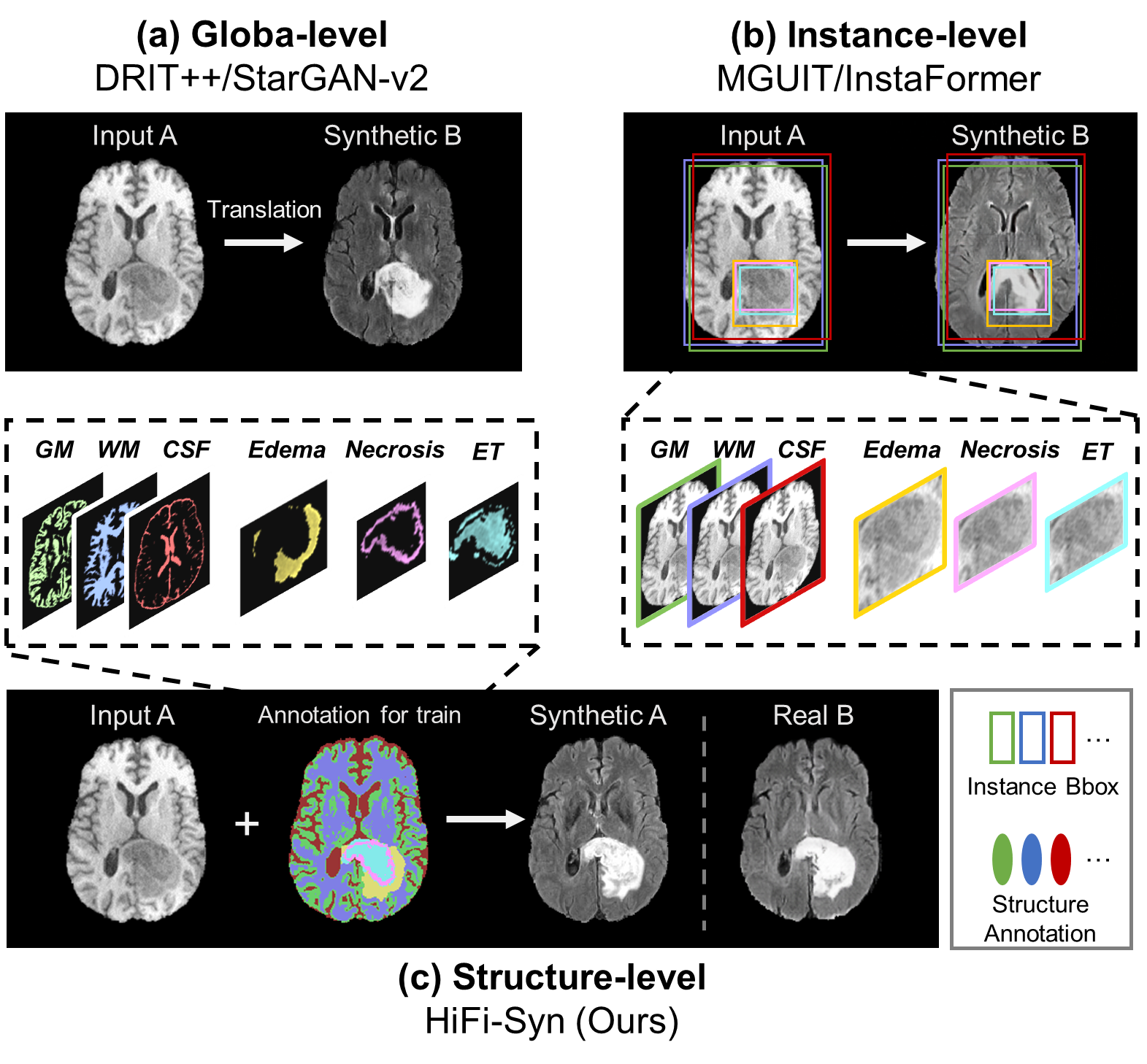} }
   \caption {Illustration of our motivation.
   Current Image-to-Image (I2I) methods, both \textbf{(a)} global-level and \textbf{(b)} instance-level, often overlook the spatial coherence of structural-level representation and the anatomical consistency of content during translation. To address this limitation, we introduce a \textbf{(c)} structure-level content translation method for synthesizing MR images containing both normal and abnormal tissues.}

   \label{fig:workflow-motivation}
\end{figure}

{I}{n} recent years, unsupervised Image-to-Image (I2I) translation has made significant progress, emerging as a promising solution for medical image synthesis in various scenarios, such as generating missing images and predicting disease outcomes~\citep{lee2021Intro2, ma2021-Intro3}. Obtaining large-scale, multi-modal magnetic resonance imaging (MRI) data is often costly and, in some cases, even impractical due to factors such as patient cooperation and medical data privacy. As an alternative, synthesizing new MRI data can provide complementary information regarding tissue morphology. Like natural image synthesis, cycle-consistent generative adversarial network (CycleGAN)~\citep{zhu2017cycleGAN} and its variants are the primary approaches for synthesizing medical images. The quality of the synthesized images is often assessed by two evaluation metrics: 1) global-level measures such as Peak Signal to Noise Ratio (PSNR); and 2) patch-level measures such as Structural Similarity Index (SSIM) and MS-SSIM. 

In most medical image synthesis tasks, the primary requirement is translation in a \textit{structure-preserving} manner. This is crucial for dependable downstream analyses and applications. Unfortunately, this requirement has not been sufficiently addressed or reflected in previous methods and evaluation metrics. Take, for example, the synthesis of brain MRI images. The resulting brain MRIs must maintain the appearance inherent to the target modality on a global scale, while also preserving the anatomical structure of the source image at a local level. Moreover, synthesizing pathological structures is of equal importance as normal anatomical structures. For instance, in T1-MRI scans, conditions like edema or brain tumors often appear as regions of decreased signal intensity. In contrast, elevated signal regions might indicate intracranial hemorrhage~\citep{saini2019comparative}.

In T2-MRI, these pathological structures may show opposite contrast distribution, with higher signals for brain tumors, and lower signals for hemorrhage.
In such scenarios, GANs are tasked with identifying the subset of pixels at the structural level for translation, which we refer to as \textit{structure-preserving translation}. However, most existing networks only model holistic contrast shifting, which aims for global-level reconstruction without the consideration of local-level structures or contexts. For example, exisiting GANs often struggle with accurately translating the signal intensity of tumors and lesions across various modalities. Moreover, these structure-wise signal changes aren't adequately captured by current metrics, compounding the challenge and more comprehensive evaluations are needed to assess the model's real capabilities.

Recently, there have been several efforts to exploit local object information in instance-level I2I translation \citep{shen2019INIT, jeong2021_MGUIT, kim2022instaformer}.
INIT~\citep{shen2019INIT} utilises a reconstruction loss in conjunction with the global translation module to independently translate the instances during training. As for testing, the model only employs the global module and neglects the instance-level details. This might lead to unexpected performance deterioration in practice. MGUIT \citep{jeong2021_MGUIT} proposes using an external memory module to store the style features for each category. However, the lack of structural and global-level guidance may lead to fragmented representations in the embedding content space. The anatomical content may also alter during translation since the training procedure has no explicit supervision. To obtain richer instance features, InstaFormer \citep{kim2022instaformer} additionally uses patches of each instance for representation learning, yet their bounding-box annotations are too coarse to capture the fine brain structures. Furthermore, in medical images, different structures are usually highly intertwined - for example, as shown in Fig.~\ref{fig:workflow-motivation}, one can refer to MR images to observe the existence of gyri and sulci, which are located on the surface of the brain in the gray matter (GM) and white matter (WM) regions. Thus, it poses challenges for the network design.

Motivated by the aforementioned issues, we aim to propose a \textit{structure-preserving} medical image-to-image translation model. Specifically, we design a disentangled and contrastive GAN-based framework that uses hierarchical granularity discrimination to exploit various level of semantic information in medical images. First, we introduce pixel-level granularity discrimination using a Brain Memory Bank (BMB). To make the memory items more discriminative and the memory space more compact, we impose constraints on them to make similar items closer and push dissimilar items away. Second, to make the translation strucure-aware, we employ structure-level granularity discrimination on each brain structure, with a re-weighting strategy to allow the network to focus on hard samples. Finally, we utilize global-level granularity discrimination to ensure the consistency of anatomical content during translation. Our experiments on three public datasets, BraTS~\citep{menze2014multimodal}, IXI\footnote{https://brain-development.org/ixi-dataset/}, and UKB~\citep{miller2016UKB}, demonstrate the superiority of our proposed framework by showcasing its ability to maintain the integrity of normal brain structures while faithfully reconstructing pathological lesions. More importantly, beyond PSNR and SSIM, we perform quantitative evaluations on related downstream tasks such as voxel-level analysis. In addition, medical experts are involved to check our generated images visually, demonstrating that our method consistently preserves the structures during image translation.

Our contributions can be summarized as follows:
\begin{itemize}	
	\item We propose a novel medical image-to-image translation framework designed to preserve both normal and pathological (abnormal) structural information during the translation process.
 	\item Our proposed framework employs a disentangled and contrastive GAN-based approach that utilizes hierarchical granularity discrimination to exploit various levels of semantic information in medical images.
 	\item We conduct extensive experiments and voxel-level evaluations on three datasets, encompassing various modalities (T1, T1ce, T2, and T2-Flair) and tasks (normal tissue and tumor). The results demonstrate the robustness of our model, outperforming previous state-of-the-art algorithms, and showcase the effectiveness of our approach in practical downstream applications, including the preservation of diagnostic value and segmentation tasks.
\end{itemize}

\section{Related work}

\subsection{Image-to-Image Translation}

Generative adversarial networks and their derivatives have been extensively utilized in the I2I translation tasks. Among the most seminal papers is Pix2Pix \citep{isola2017pix2pix}, which harnessed paired images through conditional generative adversarial networks, yielding remarkable results in image-to-image translation tasks. As the collection of paired data can usually be costly and impractical, prompting the development of numerous unsupervised I2I translation algorithms.
Following the success of CycleGAN \citep{zhu2017cycleGAN}, which used the cycle-consistency constraint for training the network, substantial attempts have been made on construction strategies for encoder and decoder, such as DiscoGAN \citep{kim2017DiscoGAN} and DualGAN \citep{yi2017DualGAN}. In order to generate multiple domain outputs, MUNIT~\citep{huang2018MUNIT}, DRIT++~\citep{lee2020drit++}, and StarGAN-v2~\citep{choi2020starganV2} were proposed. 
There are many efforts to deploy such algorithms in medical images, involving modalities such as MRI, CT, and PET \citep{ozbey2023syndiff,dalmaz2022resvit, yu2021mousegan, yu2022mousegan++, chartsias2019disentangled, chartsias2019multimodal}.
However, most of them only consider global-level information during translation, which may lead to unsatisfied performance when addressing images within multiple objects, such as organs or structures.

\subsection{Instance-level Image-to-Image Translation}

To achieve instance-aware I2I translation, several attempts \citep{shen2019INIT,jeong2021_MGUIT,kim2022instaformer} have been exploited in recent years. INIT \citep{shen2019INIT} was used to translate the entire image and instances independently. However, in the test phase, it discarded the instance-level information. DUNIT~\citep{bhattacharjee2020dunit} was proposed to employ instance consistency loss for object-awareness and further train the computationally expensive detection module, which made it inflexible. MGUIT \citep{jeong2021_MGUIT} employs bounding boxes to read and write a memory module for storing class-aware features. 
Contrastive representation \citep{yu2023gclMIA, yu2023ICCVW} serves as one of foundational concept in InstaFormer \citep{kim2022instaformer}, which incorporating patch-level contrasts within a Transformer-based network, drawing inspiration from the approach introduced by \citep{park2020CUT}.

However, the aforementioned methods have inherent limitations. For instance, coarse bounding boxes may struggle to accurately indicate instances when they are overlapped or have irregular shapes. Additionally, retrieving the memory module solely through pixel-level queries neglects their region-wise spatial relationship. Unfortunately, these concerns become even more serious when dealing with medical images.

\subsection{Structure-Preserving Translation}
In the field of medical image translation, few studies introduced structural consistency to improve the synthesis of more realistic images. Hiasa et al. proposed GC-CycleGAN~\citep{R2_hiasa2018cross} to generate corresponding pelvis CT from MRI. They introduced gradient consistency loss in CycleGAN to improve the accuracy at the boundaries. Yang el al. introduced SC-CycleGAN~\citep{R2_yang2020unsupervised} for unsupervised MR-to-CT synthesis, emphasizing structural consistency using a modality-independent neighborhood descriptor in the structure-consistency loss. Their primary focus is MRI-to-CT synthesis. While CT captures brain exterior and cranial bone structures, it lacks internal brain details crucial for neuroscience analysis.

In recent studies, mutual information loss \citep{kang2023MI} has been introduced to maintain content consistency, and outer contours of the brain \citep{phan2023mask} and organs \citep{emami2021organ} have been utilized for shape consistency. However, the dependency on implicit loss or coarse shape consistency in the mentioned approaches may present challenges, particularly when handling multi-contrast MRI images. The intricate details and variations in contrast levels across different MRI sequences demand a more nuanced and adaptable approach to ensure effective structural preservation. Taking these considerations into account, we propose a model that incorporates constraints based on brain region structures. This tailored approach not only enhances structural consistency but also ensures adaptability to the diverse discrepancies within different MRI sequences.

\subsection{Disentangled Representation for Multiple Discrimination}
GAN brings strong disentanglement prior because of its hierarchical structure~\citep{kwon2021diagonal, locatello2019challengingICML}, which naturally fits our purpose. To further promote this progress, we utilize contrastive learning ~\citep{khosla2020sup_cl} in different level discrimination to improve existing methods.

Memory network ~\citep{weston2014memory} maintains a collection of feature vectors extracted from the input data. These feature vectors can be accessed by a set of queries. Many derivatives of memory modules have been proposed ~\citep{park2020learning, gong2019memorizing, jeong2021_MGUIT}, however, similar to MGUIT \citep{jeong2021_MGUIT}, most of them mainly consider pixel-level representations. To enrich the spatial coherence of learned embedding space, we incorporate structure-wise and global-level discrimination inspired from ~\citep{hu2021region-aware}. Combined with the characteristics of medical imaging data, we also impose contrastive loss~\citep{khosla2020sup_cl} at subject-level for \textit{structure-preserving} translation.

\begin{figure*}[h!]
\centering
\includegraphics[width=0.9\textwidth]{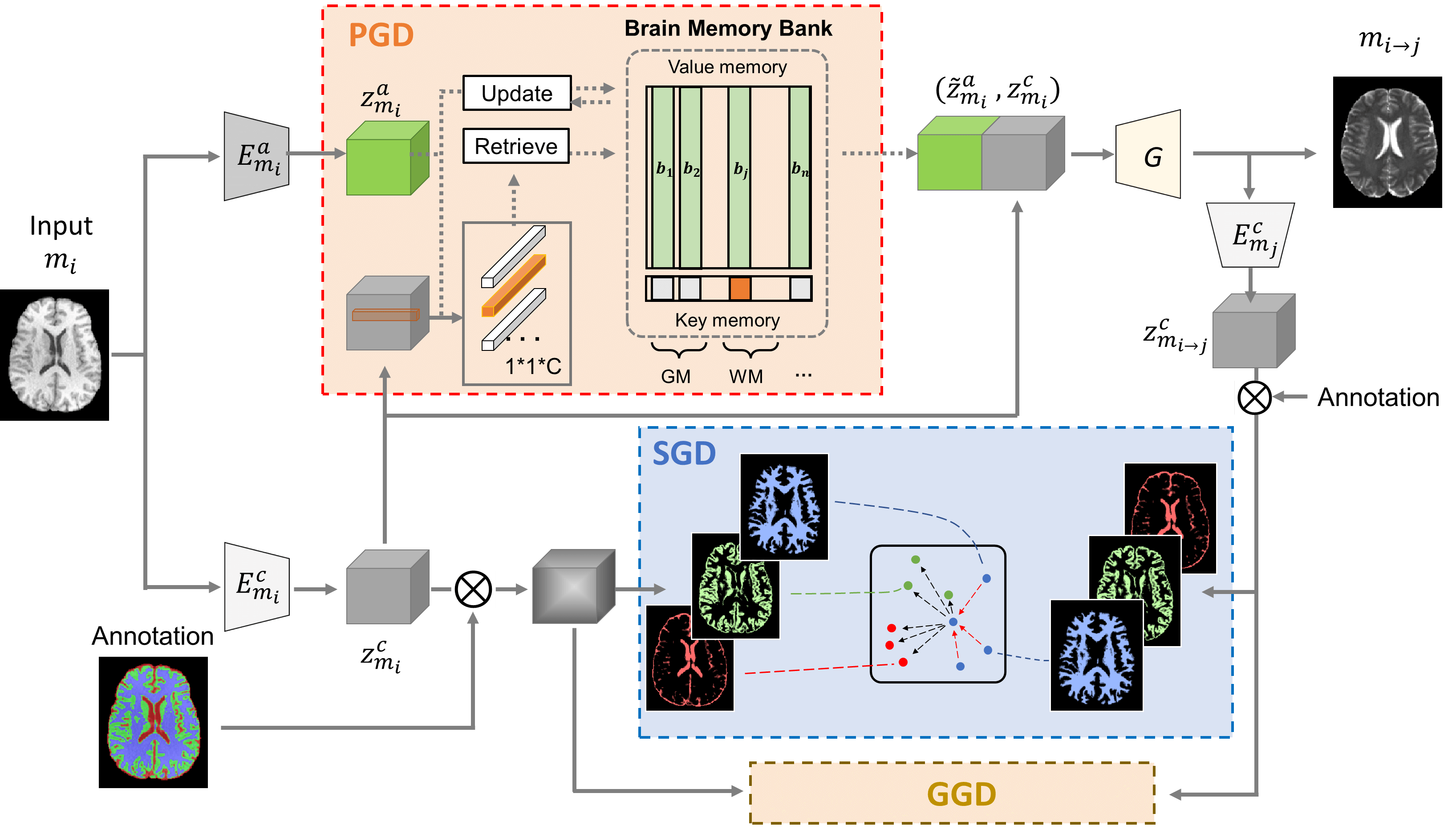}
\caption{The overview of our proposed framework. The input images are encoded to content and attribute features and then processed by Pixel-level Granularity Discrimination (PGD) module, Structure-level Granularity Discrimination (SGD) module, and Global-level Granularity Discrimination (GGD) module simultaneously to achieve structure-preserving translation. Note that the annotations are unseen during testing. For the sake of conciseness, the second-stage style translation is not illustrated here.}
\label{fig:model-architecture}
\end{figure*}

\section{Methods}

\subsection{Inductive Bias of Disentanglement}
Unsupervised disentanglement is considered impossible without the imposition of appropriate inductive biases~\citep{locatello2019challengingICML}.
In light of this, we introduce our basic assumption that medical images can be decoupled into content and attribute factors, where content factors should solely capture anatomical structural information while attribute factors should focus only on image appearance. We argue that this assumption aligns well with the fundamental nature of multi-modality MRIs. 
Fig.~\ref{fig:model-architecture} presents the proposed framework, highlighting how semantic image features are embedded into modality-specific attribute spaces and a modality-independent content space at different scales. In this subsection, we outline the basic building blocks of disentanglement inductive biases with architectural constraints. We will delve further into this assumption for I2I translation in the subsequent sections.

Considering we have $K$ number of modalities in the dataset and $\{M_i\}_{i=1}^K$ denotes each modality. $m_i$ presents each sampled image from corresponding modality. We adopt the architecture of DRIT++ \citep{lee2020drit++} as our backbone network following its success. Therefore, we omit unnecessary details to avoid repetition. In summary, our translation model includes two content encoders $ E_{m_i}^c, E_{m_j}^c $, attribute encoders $ E_{m_i}^a, E_{m_j}^a $, generators $ G_{m_i}, G_{m_j} $, and modality discriminators $ D_{m_i}, D_{m_j}$ as well as a content discriminator $D_c$. The content discriminator assists in separating attribute information from content features, facilitating disentanglement.
Given input images $ \{ m_i, m_j \} $, the first forward image style translation, we obtain encoded content features $ \{ z_{m_i}^c,z_{m_j}^c \} = \{ E_{m_i}^c(m_i),E_{m_j}^c(m_j) \} $, and attribute features $ \{ z_{m_i}^a,z_{m_j}^a \} = \{ E_{m_i}^a(m_i),E_{m_j}^a(m_j) \}$ as well as the translated pair of images $m_{i\rightarrow j}, m_{j\rightarrow i}$. Similarly, in the second style translation, we acquire features $ \{ z_{m_{i \rightarrow j}}^c,z_{m_{j \rightarrow i } }^c,  z_{m_{i \rightarrow j }}^a,z_{m_{j \rightarrow i }}^a   \} $ and reconstructed images $ \{ \hat{m}_i, \hat{m}_j   \} $ from these synthesized images. In this way, we can achieve vanilla disentanglement, yet we note that this disentanglement only takes global information into account. In the next section, we describe how we promote structure-aware disengagement via triple-level granularity discrimination.

\subsection{Structure-preserved Translation}
Hierarchical granularity analysis is crucial in medical imaging, especially in structure-preserved translation within medical analysis scenarios, as the importance of features can vary significantly across different scales. For instance, some diagnostic indicators like white matter lesions may be apparent at a local level just for a few pixels, whereas others like brain tumors need to consider a broader anatomical context. Traditional models that do not account for this hierarchical disparity might struggle with the simultaneous capture and integration of such diverse granular features due to their uniform treatment of feature importance across scales.

\subsubsection{Pixel-level Granularity Discrimination (PGD)}
\label{sec:PGD}
To ensure that the translation procedure is structure-invariant, we first introduce pixel-level granularity discrimination into our framework. Unlike the instance-granularity mechanism that treats each image as a separate entity, thus limiting in modelling a relatively global discriminative feature representation, our pixel-level granularity regards each pixel in the content encoder $ E^c $ as an individual candidate. This approach enables the capture of dense, semantically distinct feature representations. In image synthesis, the granularity of pixel-level details presents significant challenges. Pixels, being the finest elements of an image, inherently contain less structural or global information compared to larger aggregations, rendering them more sensitive to noise. This sensitivity can lead to instability during the training process and can degrade the overall quality of synthesized images. To address these issues, we introduce the Brain Memory Bank (BMB), a novel component designed to store prototypical feature representations.
The BMB is structured as $ B \in \mathbb{R}^{T \times C} $ including $ T $ slots to store value vectors with the dimension $ C $. In each slot, $ b_t $, which stands the $t_{th}$ memory item, stores domain-specific attribute representations. The corresponding key, denoted as $ k_t $, is utilized as an address to retrieve these relevant memory items. 
During the training procedure, we first retrieve the BMB and obtain enhanced domain-agnostic features $\widetilde{z}^a_{m_i}$ by using query map and then channel-wise combining the content features $z^c_{m_i}$ with the retrieved ones $\widetilde{z}^a_{m_i}$ for subsequent domain translation.

Note that, in contrast to existing memory bank techniques that only consider the content codes from the original images~\citep{jeong2021_MGUIT}, we impose contrastive consistency to consider contents of original and translated images $\{ z_{m_i}^{c} , z^{c}_{m_{i \rightarrow j }} \}$ as positive samples, thus keeping the structural information invariant at pixel-level during translation.

\noindent{\textbf{Retrieving.}}
Given the content feature $z^c_{m_i} = E^c_{m_i}(m_i)$ of size $ H \times W \times C $, where H, W, C stand for height, width, and the channel number, respectively. We define the set of pixel-wise queries $ \{ q_{m_i}^n \}_{n=1}^N $ as extracted from each pixel in the $z^c_{m_i}$
. Here, $ N= H \times W $ represents the total number of pixel-wise queries, and each query $  q_{m_i}^n $ has the size of  $ 1 \times 1 \times C $. Then, we can use softmax operation to obtain affinity scores $w_{m_i}^{n, t}$ for each key $k_t$:
\begin{equation}
w_{m_i}^{n, t} = \frac{\exp(sim ( q_{m_i}^n, k_t))}{\sum_{t^\prime=1}^{N} \exp(sim(q_{m_i}^n, k_{t^\prime} ))}
\label{eq:attention-affine-score}
\end{equation}
where $sim(\cdot, \cdot)$ denotes cosine similarity:

\begin{equation}
sim(q_{m_i}^n, k_t ) = \frac{ q_{m_i}^n \cdot {k_t}^\top }{\|q_{m_i}^n \|_2  \|  k_t  \|_2}\label{eq:cos_dis}
\end{equation}

To get the enhanced representation, $\widetilde{z}^a_{m_i}$, the memory module retrieves the memory elements that are close to $ {z}^a_{m_i}$. This procedure can be defined as follows:
\begin{equation}
\widetilde{z}^a_{m_i,n} = \sum_{t=1}^T w_{m_i}^{n,t} \cdot   b_{m_i}^t 
\label{eq:mem_raed}
\end{equation}

\noindent{\textbf{Updating.}}
For updating memory items, we choose all queries that identify the item as the close ones for each $b_i$ using the affinity scores in (1) and obtain updating weight $ u_{m_i}^k $. Then we update the keys and values in BMB through:
\begin{equation}
\widetilde{k_t} \leftarrow \|  \alpha_p \cdot k_t + (1- \alpha_p ) \cdot \sum \nolimits_{n^\prime=1}^N [ u_{m_i}^{n,t} \cdot q_{m_i}^{n^\prime} + u_{m_j}^{n,t} \cdot q_{m_j}^{n^\prime}]   \|_2 
\label{eq:update-k}
\end{equation}
\begin{equation}
 \widetilde{b}^t_{m_i} \leftarrow \|  \alpha_p \cdot b^t_{m_i} + (1- \alpha_p ) \cdot \sum \nolimits_{n^\prime=1}^N u_{m_i}^{n,t} \cdot \widetilde{z}^a_{m_i,{n^\prime}}   \|_2 
\label{eq:update-b}
\end{equation}
\begin{equation}
u_{m_i}^{n,t} = \frac{\exp(sim ( q_{m_i}^n, k_t))}{\sum_{n^\prime=1}^{N} \exp(sim(q_{m_i}^{n^\prime}, k_t ))}
\label{eq:update-u}
\end{equation}
where $\alpha_p $ is to control updating rate. Similarly, we can compute $u_{m_j}^{n,t}$ and $\widetilde{b}_{m_j}^t$ for items stored attributes from other domains.
The keys and values within the BMB are updated using a momentum-based approach, employing a weighted blend of existing values and aggregated inputs from related queries. The update rate is controlled by \( \alpha_p \). This updating strategy stabilizes the pixel-level representation learning process by preventing abrupt changes in the pixels. This controlled updating process helps in maintaining the robustness of the system against noise and outlier data, which might otherwise lead to significant challenges in learning. The gradual integration of new information ensures that the BMB develops a more comprehensive  understanding of the data patterns, leading to more reliable and consistent outputs during translation.

Furthermore, by symmetrically updating across different modalities, the BMB enhances cross-domain discriminative capabilities at the pixel level, which helps to further decouple the content and attributes of each modality. This is particularly beneficial in scenarios where attributes from one domain can inform or enhance understanding in another. For instance, in multi-modal medical imaging, understanding the differences and relationships between T1 and T2 images can significantly improve the accuracy and fidelity of generated images.

To promote learning a better-disentangled feature representation, we introduce a contrastive prior here for each specific subject, the content features of input image $m_i$ and translated images $m_{i \rightarrow j}$ are close to each other in latent content space since they share the same intrinsic structures (positive pairs, $\Omega_p^+ $). In contrast, images of different subjects (even if they belong to the same modality) are pushed away. Based on this assumption, we impose pixel-level contrastive loss, which could be formulated as follows:
\begin{equation}
{{\cal L}_{PGD} = \sum_{i=1}^N \frac{-1}{| \Omega^+_{p} |} \sum_{m^+ \in \Omega^+_{p} } \log \frac{{\exp \left( {  CL^+ /{\tau_1 }} \right)}}{{ \sum_{m \in \Omega_{p}^{'}} {\exp \left( {  CL /{\tau _1 }} \right)} }}}
\label{eq997}
\end{equation}
where $ CL^ + = sim ( q_{ m_i}^n , q_{m^+}^n ) $, $ CL =  sim ( q_{ m_i}^n ,  q_{ m}^n  ) $ and $ \Omega_{p}^{'} = \Omega_{p} ~  \backslash \{ q_{m_i}^n \} $. $\tau_{1}$ is the temperature scaling parameter.

To mitigate the decline in performance during the test phase when structural annotations are not accessible, we introduce a global-wise memory items to BMB. For symmetry, the number of global-wise items is equal to the sum of structure-wise items.

\subsubsection{Structure-level Granularity Discrimination (SGD)}
In the above pixel-level granularity discrimination module, each pixel in the content space is treated independently of the other. Here we go one step further to explore structure-level granularity discrimination (SGD), thereby calibrating the deformation of structural semantic content during translation. By leveraging available training data with annotated brain structures, SGD can learn decoupled features at the structure-level, thus capturing more characteristic variations between different anatomical structures in the brain. We show in our experiments later that SGD is a critical intermediate component that leads to more efficient and robustness learning than existing methods, either learning from the pixel-level or global level.

Similarly, we use contrastive learning in SGD to pull region semantic features of the same structures together while keeping features of different structures apart. Thereby, we reuse the disentangled content features $ z_{m_i} \in \mathbb{R}^{H \times W \times C} $ and annotated structural ground truth $ G_{m_i} \in \mathbb{R}^{H \times W} $ which are down-sampled to the $H, W$ matrix size. Denoting $ L $ as the number of structures $ \{ s_i \}_{i=1}^L $, with the guidance of $G_{m_i}  $, we can obtain structure-wise content $ z^{c, s_i}_{m_i} $ of modality $ M_i $ via:
\begin{equation}
z^{c, s_i}_{m_i} = z_{m_i}(x, y) \odot \mathds{1} \left [  G_{m_i}(x,y) = s_i \right ] 
\label{eq995}
\end{equation}
where $\odot $ presents Hadamard product and $(x,y)$ is position coordinate. $\mathds{1}(\cdot) $ is the binary indicator denoting  whether the pixel belongs to $ s_i$. Similarly, we have translated $z^{c, s_i}_{m_{i \rightarrow j}} $  of modality $ M_{i \rightarrow j } $. Though they can be assigned to diverse appearance conditioned by different attribute codes, the anatomical content of $\{ z^{c, s_i}_{m_i}, z^{c, s_i}_{m_{i \rightarrow j}} \}$ from the same subject ought to be accordant which belongs to positive samples $ \{ z^{c, s_i}_{m^+ } \} \in \Omega^+_{s_i} $. Negative samples $  \Omega^-_{s_i} $ include residual structures and anatomical content features from other subjects. Moreover, to force the network to pay more attention to hard examples, we design a content similarity-based weighting scheme to automatically adjust weights for each sample. Specifically, given a set of structures $\{ s \}$ from one subject regardless of the modality, the deformation maps $ \{ d_{s_i} \}$ guide the weight between each two positive samples, and the ranking-based weighting calculation can be defined as:
\begin{equation}
d_{s_i} = \sum \limits_{x,y}  \left|  z^{c, s_i}_{m_i} - z^{c, s_i}_{m_{i \rightarrow j}} \right|
\label{eq998}
\end{equation}

\begin{equation}
w_{s_i} = \exp ( -\alpha_s \cdot \text{rank}(d_{s_i}) ) 
\label{eq999}
\end{equation}
where hyper-parameter $\alpha_s $ controls the degree of smoothness in the exponential function, and $ \text{rank} (\cdot)$ involves sorting the values in a set. $ w_{s_i} $ varies from 0 to 1. Pairs with greater deformation are accorded larger weights. It can harness the network to correct mismatches, thereby preserving the brain structure during the style translation. Finally, the overall loss for structure-level granularity discrimination can be defined as follows:
\begin{equation}
{{\cal L}_{SGD} = \sum_{i=1}^N \frac{-1}{| \Omega^+_{s} |} \sum_{m^+ \in \Omega^+_{s} } \log \frac{{\exp \left( { w_{s_i} \cdot CL^+ /{\tau_2 }} \right)}}{{ \sum_{m \in \Omega_{s}^{'}} {\exp \left( {  CL /{\tau _2 }} \right)} }}}
\label{eq:SGD}
\end{equation}
where $ CL^+ = sim( z^{c, s_i}_{m_i}, z^{c, s_i}_{m^+ }  ) $, $ CL = sim( z^{c, s_i}_{m_i}, z^{c, s_i}_{m}  ) $, $m^+ \in \Omega^+_{s} $, and $ \Omega_{s}^{'} = \Omega_{s} \backslash \{ s_i \}$. $ \tau_2 $ is the temperature scaling parameter

\subsubsection{Global-level Granularity Discrimination (GGD)}

Similar to the previous SGD, the global anatomical content from the same entity should remain consistent, regardless of the target modalities to which it is translated. To ensure this consistency, we leverage the entire content feature $ z^c_{m_i}$ and apply a contrastive learning loss, which is similar to the equation~(\ref{eq:SGD}) but with $w_{s_i}=1$. Content features from identical subjects constitute positive pairs and are denoted by $ \Omega_{c}^+ $. Conversely, features from disparate subjects are considered negative pairs. The formulation is as follows:

\begin{equation}
{{\cal L}_{GGD} = \sum_{i=1}^N \frac{-1}{| \Omega^+_{g} |} \sum_{m^+ \in \Omega^+_{g} } \log \frac{{\exp \left( {  CL^+ /{\tau_3 }} \right)}}{{ \sum_{m \in \Omega_{g}^{'}} {\exp \left( {  CL /{\tau _3 }} \right)} }}}
\label{eq:GGD}
\end{equation}
where $ CL^+ = sim( z^{c}_{m_i}, z^{c}_{m^+ }  ) $, $ CL = sim( z^{c}_{m_i}, z^{c}_{m}  ) $, $m^+ \in \Omega^+_{s} $, and $ \Omega_{c}^{'} = \Omega_{c} \backslash \{ m_i \}$. $ \tau_3 $ is the temperature scaling parameter.

\subsection{The Overall Loss Function}

Following successful practices \citep{lee2020drit++, choi2020starganV2}, we used the following losses to train the model.

\noindent{\textbf{Reconstruction loss.}}
The bidirectional reconstruction losses encourage disentanglement by updating the gradient propagation among blocks in a synergistic manner. We denote cross-cycle and self-reconstruction as $ {\cal L}_1^{cyc}$ and $ {\cal L}_1^{self}$, respectively. We formulate these losses as follows: 

\begin{equation}
\begin{aligned}
\hspace {-0pc}
{\cal L}_1^{cyc} ( E_{m}^c,E_{m}^a,{G_{m}}  ) = 
\Vert \hat{m}_i - m_i  {\Vert _1}
+ \Vert \hat{m}_j - m_j   {\Vert _1}
\end{aligned}
\end{equation}

\begin{equation}
\begin{aligned}
& \hspace {-0.5pc}  {\cal L}_1^{self-recon} \left( {E_{m}^c,E_{m}^a,{G_{m}}} \right)  = \\& \Vert {G_{m_i}} ( z_{m_i}^c, z_{m_i}^a ) - {m_i}{\Vert _1}  +  \Vert {G_{m_j}} ( z_{m_j}^c, z_{m_j}^a ) - {m_j}{\Vert _1}
\end{aligned}
\end{equation}

\noindent{\textbf{Adversarial loss.}}
$ {\cal L}_{adv}^{cont}$ and ${\cal L}_{adv}^{domain} $ are used for adversarial learning. The former loss guarantees that that image modality cannot be inferred from content codes alone, aiding in the disentanglement of attribute information from content features. The later ensures the synthetic images has the modality-specific features as the actual ones. These losses can be formulated as follow:
\begin{equation}  
\begin{aligned}
& \hspace {-2pc} 
{\mathop {\min }\limits_{\left( {E^{c},{G}} \right)} \mathop {\max }\limits_{D^c} {\cal L}_{adv}^{cont}} =
\\& \hspace {-0.7pc}  \mathbb {E}_{m_i}[ \frac{1}{2}  \text {log}D^c ( z^c_{m_i} ) +  \frac{1}{2}  \text {log}(1 - D^c ( z^c_{m_i} )) ] + \\& \hspace {-0.7pc} \mathbb {E}_{m_j}[ \frac{1}{2}  \text {log}D^c ( z^c_{m_j} ) +  \frac{1}{2}  \text {log}(1 - D^c ( z^c_{m_j} )) ]
\end{aligned}
\end{equation}
\begin{equation}
\begin{aligned}
& \hspace {-1pc} 
{\mathop {\min }\limits_{\left( {E^{c},E^{a},{G}} \right)} \mathop {\max }\limits_{D} {\cal L}_{adv}^{domain}} =
\\& \mathbb {E}_{m_i}[ \text {log}D_{m_i} (m_i) ] +  \mathbb {E}_{m_{j\rightarrow i}}[ \text {log}(1- D_{m_i} (m_{j\rightarrow i})+ \\& \mathbb {E}_{m_j}[ \text {log}D_{m_j} (m_j) ) ]   + \mathbb {E}_{m_{i\rightarrow j}}[ \text {log}(1- D_{m_j} (m_{i\rightarrow j})]
\end{aligned}
\end{equation}

\noindent{\textbf{Overall loss.}}
The full loss function of our framework is:
\begin{equation}  
\begin{aligned}
& \hspace {-1.0pc} 
{\mathop {\min }\limits_{\left( {E^{c},E^{a},{G}} \right)} \mathop {\max }\limits_{\left(D, D^c\right) } {\cal L}(E^c,E^a,G}) =
\\& \hspace {-0.7pc}  \lambda_{1} {\cal L}_{adv}^{cont} + \lambda_{2} {\cal L}_{adv}^{domain} + \lambda_{3} {\cal L}_{1}^{cyc} + \lambda_{4} {\cal L}_{1}^{self-recon} +
\\& \hspace {-0.7pc} \lambda_{5} {\cal L}_{PGD} + \lambda_{6} {\cal L}_{SGD} + \lambda_{7} {\cal L}_{GGD}
\end{aligned}
\end{equation}

\begin{figure}[h!]
  \centering
\includegraphics[width=0.9\linewidth]{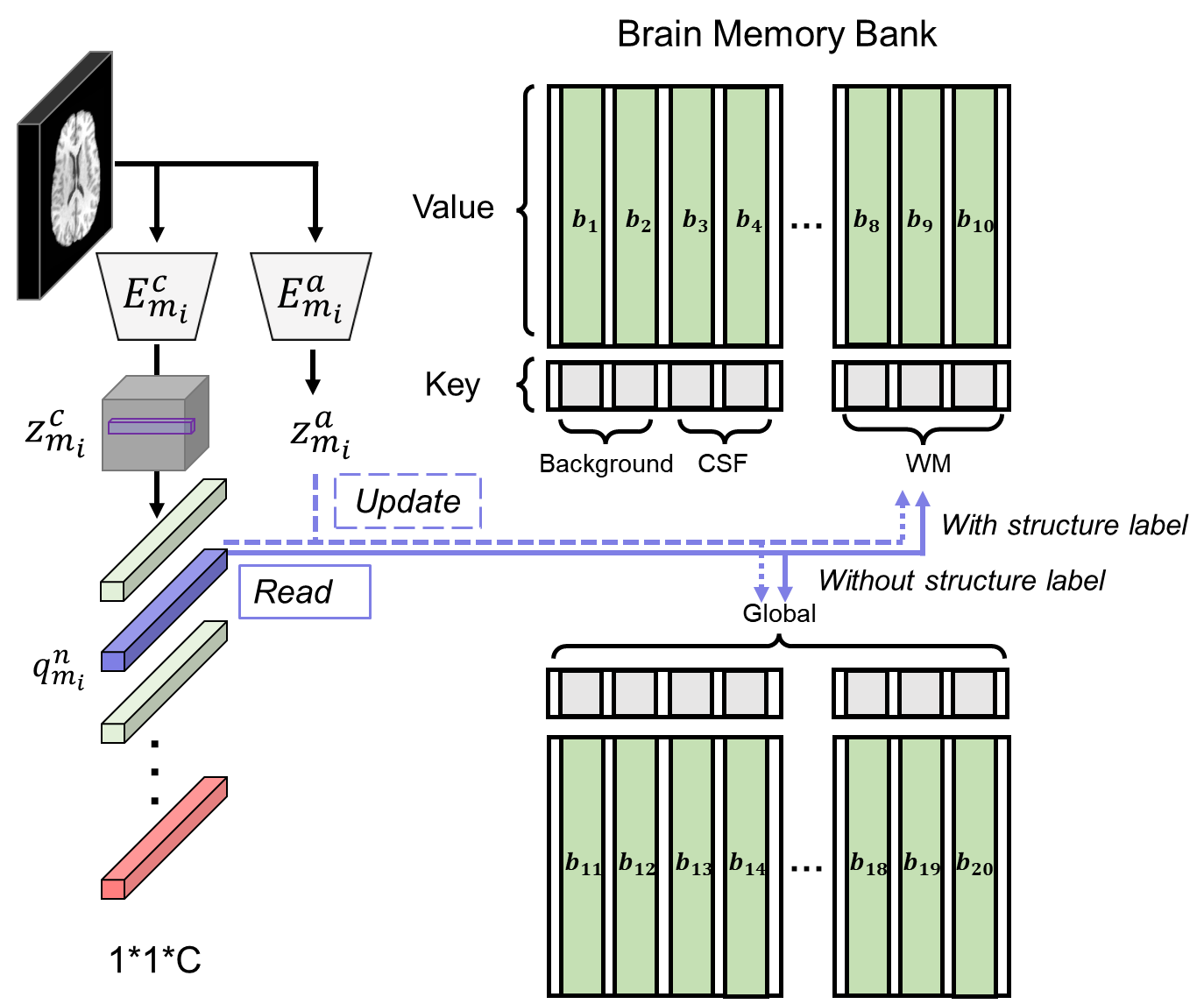}

   \caption{{Illustration of memory item setting for the IXI/UKB datasets.}
   Each query reads and updates the corresponding structure-aware items with labels and global items without labels during training. Items store domain-specific attribute representations, utilizing shared keys to access them.
   }
   \label{fig:bank-details}
\end{figure}

\section{Experimental Setup}
\subsection{Datasets}
We conducted experiments on three widely recognized public datasets of brain MRIs. 

\subsubsection{BraTS Dataset} The 2018 Multimodal Brain Tumor Segmentation Challenge (BraTS) dataset \citep{menze2014multimodal} consists of 285 annotated MRI subjects from glioma patients across multi-institution. Each subject has four aligned modalities, T1, T1ce, T2, and FLAIR. The MRI volumes had been skull-stripped and re-sampled to the resolution of $1\times1\times1~mm^3$. We used 100 subjects for I2I translation, where 70 are reserved for training and 30 are for testing. From each subject, 100 axial cross-sections containing brain tissues are selected. This dataset contains 3 classes of tumor labels (core, enhancing, and whole tumor) initially. Further, we invite two expert radiologists to delineate and curate the segmentation of gray matter (GM), white matter (WM), and cerebrospinal fluid (CSF) using FSL~\citep{jenkinson2012fsl} and itk-snap~\citep{yushkevich2016itk}. Finally, we have six classes in total, which include GM, WM, and CSF, as well as edema, necrosis, and enhancing tumor regions.

\subsubsection{IXI Dataset} In IXI dataset, T1 and T2 brain MR images from 80 healthy subjects are used in our study, where 60 are reserved for training and 20 for testing. Like in BraTS, we also select 100 axial cross-sections containing brain tissues for IXI dataset. Note that MR images from the IXI dataset are unregistered, so we employed BEN \citep{yu2022BEN} and ANTs \citep{avants2009ANTs} to register them at subject-level after skull-stripping to get real reference images for evaluating I2I translation performance on the IXI dataset. Besides, no segmentation labels were provided in this dataset. Therefore, we ask two expert radiologists to provide us with three classes of GM, WM, and CSF, with the assistance of FSL~\citep{jenkinson2012fsl} and itk-snap~\citep{yushkevich2016itk}.

\subsubsection{UKB Dataset} This dataset comprises MR images sourced from UK Biobank \citep{miller2016UKB}. In our study, we employed T1 and T2-Flair MRI images from 100 healthy subjects, with 70 individuals allocated for training and 30 for testing. Consistent with the setting for the BraTS and IXI datasets, we select 100 axial cross-sections containing brain tissues for our analysis and obtain structural labels in this dataset with those criteria used in the IXI dataset. Incorporating the UKB dataset into our study enhances the diversity and comprehensiveness of our datasets and experiments, as it encompasses a broader spectrum of subjects and imaging scenarios commonly encountered in real-world clinical practice.

\subsection{Implementation Details}
Our model is implemented using Pytorch~\citep{imambi2021pytorch}, building upon the implementation of DRIT++~\citep{lee2018diverse}, and trained on one NVIDIA Tesla V100 GPU. Input images are resized to $256 \times 256$ matrix size and are normalized to the range of [-1, 1]. We use Adam optimizer with a learning rate of 0.0001, and set the size of the batch to 2. We set the content features $ z^c_{m_i} $ with height $H$, weight $W$, and feature channel number $C$ to 54, 54, and 256, respectively.

In the proposed architecture, the discriminator $D_{m_i}$ is designed to evaluate the authenticity of an input by processing it through a series of convolutional layers. $D_{m_i}$ takes ${m_i}$ as input, which is first processed by a convolution layer with a 3x3 kernel and 64 output channels. The output is then passed through sequential layers with 3x3 kernels, and the number of output channels is increased progressively, moving from 128 to 256, 512, and 1024. Then, followed by two layers contain convolution operations with a 3x3 kernel and 1024 output channels. Finally, a 1x1 convolution layer produces a single output, representing a binary classification, indicating whether the input is real or generated.
The content discriminator $D_{c}$ receives $z^c $ as input, starting with a convolution layer using three 7x7 kernels with 256 output channels, followed by a 4x4 kernel layer. The final layer is a 1x1 convolution that produces a single output for binary classification. $D_{c}$ is designed to distinguish extracted content representations between two modalities, promoting disentangled representation learning.

The Brain Memory Bank (BMB) is divided into two distinct parts. The items belonging to the pixel-level granularity discrimination (PGD) module solely focus on structural-wise knowledge, whereas the remaining items encompass global-level information, attributed to global-level granularity discrimination (GGD). Empirically, these two parts are symmetrical in size, each containing an equal number of items.
In accordance with this symmetrical configuration, the number of memory items for the IXI/UKB dataset is set to 20, while for the BraTS dataset, it is set to 38 items. In instance, for the IXI dataset, the 20 memory items are divided into 2 for the background, 2 for CSF, 3 for GM, 3 for WM, and 10 for global structures.
The diagram is presented in Fig.~\ref{fig:bank-details}. In contrast to MGUIT~\citep{jeong2021_MGUIT}, we employ structure-wise annotations instead of bounding boxes to locate pixel positions, resulting in more precise structural descriptions. Additionally, we introduce items to record global values in the BMB, recognizing that annotations are not available during the testing phase. We expect the network to learn contextual discrepancies from the input rather than solely relying on labels, thereby ensuring the robustness and generalizability of the network. 

The hyperparameter $\{ \lambda_{1-4} \}$ are set the same as DRIT++~\citep{lee2020drit++}. Empirically, we set $\alpha_p = 0.01$,  $\alpha_s =0.5$, $\lambda_5,\lambda_7=1 $, and $\lambda_6=2$.

\subsection{Evaluation Metrics}
We trained all I2I translation networks in an unpaired manner and evaluated their performance using the Peak Signal Noise Rate (PSNR) and Structural Similarity Index Measure (SSIM). In addition, we incorporated the Dice score and volumetric similarity metrics to evaluate the anatomical-level quality of image translation. By focusing on segmented structures of interest, we are able to achieve a more reliable and comprehensive measurement of structural fidelity during image synthesis.

\subsection{Competitive Methods}
We compare our proposed framework with unsupervised image-to-image translation methods, including CycleGAN \citep{zhu2017cycleGAN}, AttentionGAN \citep{tang2021attentiongan}, MUNIT \citep{huang2018MUNIT}, DRIT++ \citep{lee2020drit++},  UVCGAN \citep{torbunov2023uvcgan}, StarGAN-v2 \citep{choi2020starganV2}, as well as the recently state-of-the-art (SOTA) instance-level I2I methods MGUIT \citep{jeong2021_MGUIT} and InstaFormer \citep{kim2022instaformer}. In addition, four medical structure-constrained translation methods are presented to demonstrate the effectiveness of our proposed approach. These methods include GC-CycleGAN~\citep{R2_hiasa2018cross}, SC-CycleGAN~\citep{R2_yang2020unsupervised}, COSMOS \citep{shin2022cosmos}, and RegGAN \citep{kong2021RegGAN}.

\begin{itemize}	
    \item \textbf{CycleGAN:} A classic unsupervised image-to-image translation method that has been widely adopted in various domains. It employs a cycle-consistency loss to ensure consistency between the translated and original images in both directions, contributing to improved image quality and preservation of content.
    \item \textbf{GC-CycleGAN:} It introduces a gradient consistency loss into the traditional CycleGAN architecture. The gradient consistency loss is employed to improve the accuracy of the generated images, particularly focusing on addressing challenges at the boundaries.
    \item \textbf{SC-CycleGAN:} This method is developed for unsupervised image synthesis, with a primary focus on preserving structural consistency between the translated images. It introduces a modality-independent neighborhood descriptor within the structure-consistency loss to emphasize the importance of structural details.
    \item \textbf{COSMOS:} This method won first place in the Cross-Modality Domain Adaptation (CrossMoDA) challenge (https://crossmoda-challenge.ml). COSMOS performs translation and segmentation tasks simultaneously, enabling accurate identification of key structural components in the translated images.
    \item \textbf{AttentionGAN:} This approach integrates attention mechanisms into the traditional CycleGAN framework, allowing the model to focus on critical regions of the image during translation. By incorporating attention layers, AttentionGAN enhances translation quality by selectively emphasizing relevant areas while reducing noise in other regions.
    \item \textbf{MUNIT:} MUNIT focuses on multimodal translation tasks, aiming to generate diverse outputs from a single input image. It introduces disentangled representations for content and style, enabling more flexible and controlled image synthesis across different domains.
    \item \textbf{DRIT++:} DRIT++ builds upon the disentangled representation framework, aiming for diverse image translation. It introduces additional enhancements to improve the diversity and quality of the generated images while maintaining disentanglement for better control over the translation process.
    \item \textbf{RegGAN:} This method is designed to address the challenges of medical imaging, where paired, pixel-wise aligned data are not always available. RegGAN considers misaligned target images as noisy labels. By integrating a registration network, RegGAN aims to adaptively fit the misaligned noise distribution.
    \item \textbf{UVCGAN:} It enhances the classic CycleGAN architecture by integrating a Vision Transformer (ViT) within the generator. To ensure robustness and training stability, UVCGAN incorporates additional training and regularization techniques, such as gradient penalties and self-supervised pre-training.
    \item \textbf{StarGAN-v2:} StarGAN-v2 extends the capabilities of its predecessor by enabling image translation across multiple domains with a single model. It introduces a novel method for learning domain-specific style codes, resulting in improved image quality and diversity in the translated outputs.
    \item  \textbf{MGUIT:} It introduces a memory mechanism to store and utilize relevant information during the translation process. By incorporating memory guidance, MGUIT aims to enhance the model's ability to generate accurate and contextually relevant translations.
    \item \textbf{InstaFormer:} InstaFormer is a recent state-of-the-art instance-level image-to-image translation method. It leverages a transformer-based architecture to capture long-range dependencies and context information, enhancing the model's ability to generate realistic and coherent translations at the instance level.

\end{itemize}

\begin{figure*}[h!] 
  \centering
  \includegraphics[width=0.8\linewidth]{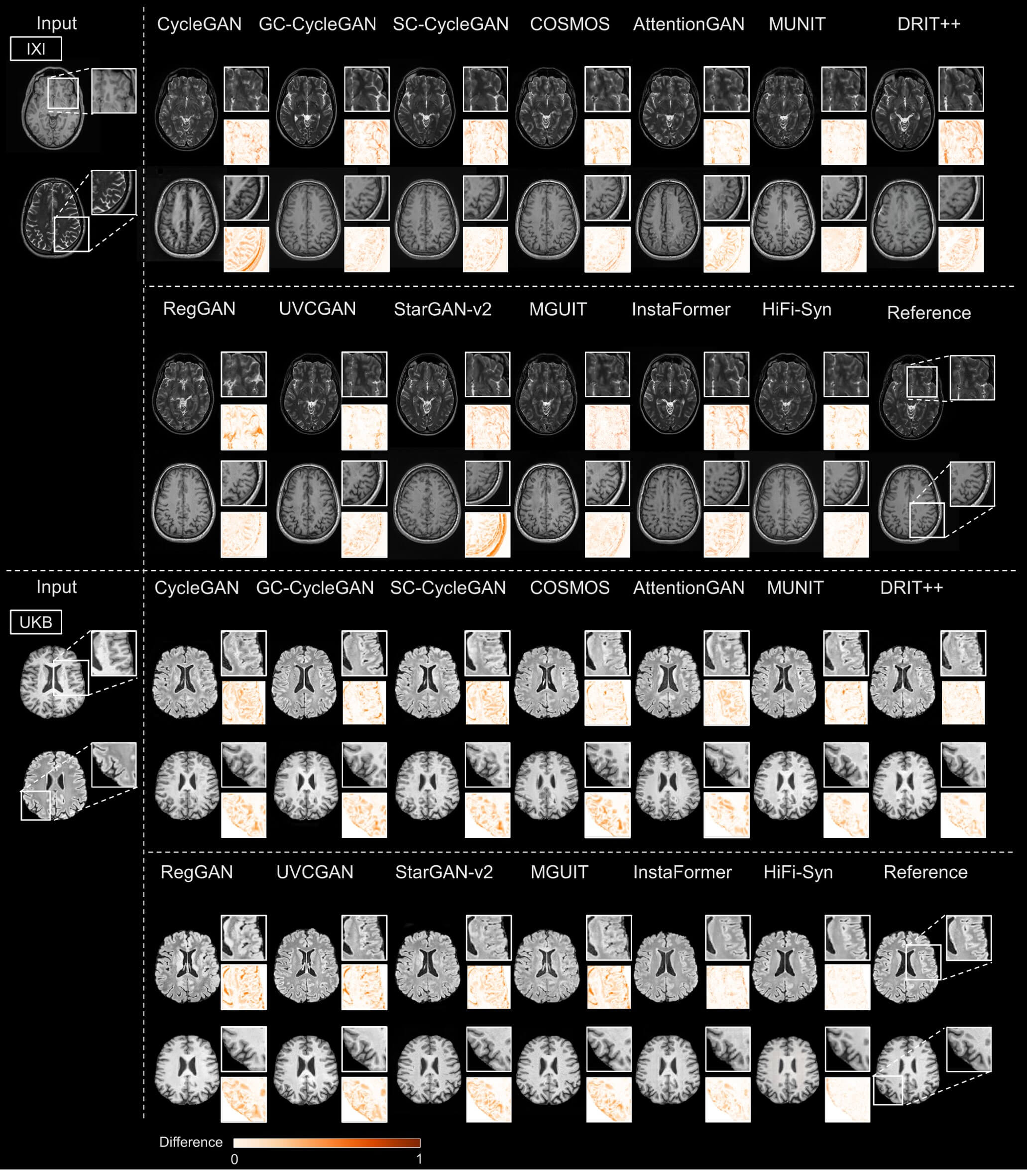}
  \vspace{-0.10in}  
   \caption{{Comparison of I2I translation methods on the IXI and UKB datasets.} (From the first to the last sample) T1$\rightarrow$T2, T2$\rightarrow$T1, T1$\rightarrow$T2-Flair, and T2-Flair$\rightarrow$T1 results. Compared to other SOTA methods, HiFi-Syn shows superior structure-preserving translation with clearer error maps.}
   \label{fig:Comparison-sota-results}
  \vspace{-0.12in}  
\end{figure*}

\section{Results and Discussion}

\subsection{Quantitative Evaluation on Normal Brain Samples}
We initiate our experiments with a quantitative evaluation of our proposed structure-preserving synthesis technique on normal brain images, using two datasets: IXI and UKB.
By examining the results of these experiments, we aim to quantify its effectiveness in faithfully reproducing the intricate structural details present in normal brain images.

Fig.~\ref{fig:Comparison-sota-results} shows several exemplary synthetic images of the SOTA methods on the IXI and UKB datasets. It can be observed that for both datasets, InstaFormer and StarGAN-v2 struggle to preserve the anatomical structures of gyri and sulci. This becomes particularly evident in the second example, as illustrated in Fig.~\ref{fig:Comparison-sota-results}. Although DRIT++ and MGUIT shows distinct contrast and boundaries, they are peculiarly prone to generate unrealistic anatomies in the T1$\leftrightarrow$T2-Flair tasks. This is unsurprising, while they have decoupled the MR images into attribute and content spaces, they lack structure-wise guidance during translation. This absence can result in fragmented and ambiguous embedding spaces and is insufficient to ensure the preservation of fine brain structure during translation. 
Thanks to our proposed hierarchical triple-level granularity discrimination, our synthetic images preserve structures during translation and reproduce the anatomic-specific contrast variations between multi-modality MRI. We observe the successful preservation of critical anatomical landmarks, including ventricles, sulci, and gyri. The quantitative results of PSNR and SSIM in Table \ref{tb:comparision-sota-ixiukb} confirm the superiority of HiFi-Syn, which surpasses other methods in most translation tasks.

\begin{table*}[h!]   
\centering
\caption{Quantitative evaluation on the IXI and UKB datasets. We perform bidirectional translation for each domain pair and evaluate the performance using PSNR and SSIM metrics, where higher values indicate better results. The p-value is obtained from the Wilcoxon Signed-Rank Test results.}
\vspace{+0.2cm}
\resizebox{1.0\textwidth}{!}{
\begin{tabular}{l|cc cc|cc|cc|cc cc|cc|cc}
\hline
\multirow{3}{*}{Method} & \multicolumn{8}{c|}{IXI} & \multicolumn{8}{c}{UKB} \\ 
                        \cline{2-17}  
                        & \multicolumn{2}{c}{T1 $\rightarrow$ T2} & \multicolumn{2}{c|}{T2 $\rightarrow$ T1} & \multicolumn{2}{c|}{Avg} & \multicolumn{2}{c|}{P-value} & \multicolumn{2}{c}{T1 $\rightarrow$ T2} & \multicolumn{2}{c|}{T2 $\rightarrow$ T1} & \multicolumn{2}{c|}{Avg} & \multicolumn{2}{c}{P-value} \\ 
                        \cline{2-17}  
                        & PSNR & SSIM  & PSNR & SSIM & PSNR & SSIM & PSNR & SSIM & PSNR & SSIM & PSNR & SSIM & PSNR & SSIM  & PSNR & SSIM \\ \hline \hline
CycleGAN                & 19.7149 & 0.7739 & 20.3284 & 0.7862 & 20.0217 & 0.7801 & 6e-72 & 9e-65& 21.8037 & 0.7404 & 22.2846 & 0.7718 & 22.0442 & 0.7561 & 4e-48 & 2e-50 \\
GC-CycleGAN             & 19.7751 & 0.7762 & 20.3541 & 0.7914 & 20.0646 & 0.7838 & 1e-45 & 2e-46& 21.8138 & 0.7527 & 22.4855 & 0.7864 & 22.1497 & 0.7696 & 6e-29 & 1e-42 \\
SC-CycleGAN             & 20.6816 & 0.7852 & 20.4581 & 0.8003 & 20.5699 & 0.7928 & 4e-77 & 7e-20& 22.0411 & 0.7648 & 22.5284 & 0.7903 & 22.2848 & 0.7776 & 1e-36 & 5e-29 \\
COSMOS                  & 20.7713 & 0.7909 & 20.8709 & 0.8122 & 20.8211 & 0.8016 & 4e-42 & 8e-46& 21.7288 & 0.7571 & 22.9679 & 0.7976 & 22.3484 & 0.7774 & 3e-42 & 8e-52 \\
AttentionGAN            & 21.8861 & 0.8027 & 21.4457 & 0.8173 & 21.6659 & 0.8100 & 8e-56 & 8e-53& 22.3332 & 0.7798 & 23.4061 & 0.8103 & 22.8697 & 0.7951 & 4e-56 & 4e-36 \\
MUNIT                   & 20.9800 & 0.7833 & 21.7122 & 0.8024 & 21.3461 & 0.7929 & 4e-30 & 9e-21& 22.3803 & 0.8074 & 23.4913 & 0.8106 & 22.9358 & 0.8090 & 9e-70 & 3e-46 \\
DRIT++                  & 22.3210 & 0.8050 & 22.2411 & 0.7722 & 22.2810 & 0.7886 & 6e-31 & 2e-12& 22.4867 & 0.8115 & 23.3427 & 0.8261 & 22.9147 & 0.8188 & 2e-57 & 8e-62 \\
RegGAN                  & 21.3543 & 0.7937 & 21.8756 & 0.8144 & 21.6150 & 0.8041 & 3e-25 & 6e-41& 22.9104 & 0.8133 & 23.6335 & 0.8194 & 23.2720 & 0.8164 & 4e-40 & 3e-37 \\
UVCGAN                  & 22.0915 & 0.8048 & 22.1106 & 0.8193 & 22.1011 & 0.8121 & 5e-37 & 9e-30& 23.1032 & 0.8209 & 23.8974 & 0.8245 & 23.5003 & 0.8227 & 4e-34 & 1e-21 \\
StarGAN-v2              & 22.1558 & 0.7975 & 22.5482 & 0.8131 & 22.8520 & 0.8053 & 8e-24 & 5e-42& 22.9423 & 0.8158 & 23.7714 & 0.8305 & 23.3569 & 0.8232 & 4e-12 & 1e-24 \\
MGUIT                   & \textbf{24.3749} & 0.7949 & 23.2653 & 0.8231 & 23.8201 & 0.8090 & 7e-21 & 4e-64& 23.2816 & 0.8279 & 24.5478 & 0.8351 & 23.9147 & 0.8315 & 8e-13 & 5e-14 \\
InstaFormer             & 22.3974 & 0.8108 & 22.8315 & 0.8259 & 22.6145 & 0.8184 & 9e-18 & 1e-10 & 23.1576 & 0.8290 & \textbf{25.0543} & 0.8372 & 24.1060 & 0.8331 & 1e-25 & 6e-4 \\
HiFi-Syn                & 23.8376 & \textbf{0.8240} & \textbf{24.1751} & \textbf{0.8371} & \textbf{24.0064} & \textbf{0.8306} & -       & -       & \textbf{23.3611} & \textbf{0.8337} & 24.9759 & \textbf{0.8459} & \textbf{24.1685} & \textbf{0.8398} & -       & -       \\

\hline
\end{tabular}}
\label{tb:comparision-sota-ixiukb}
\end{table*}

\begin{figure*}[h!]
	\centering
	\includegraphics[width=0.85\textwidth]{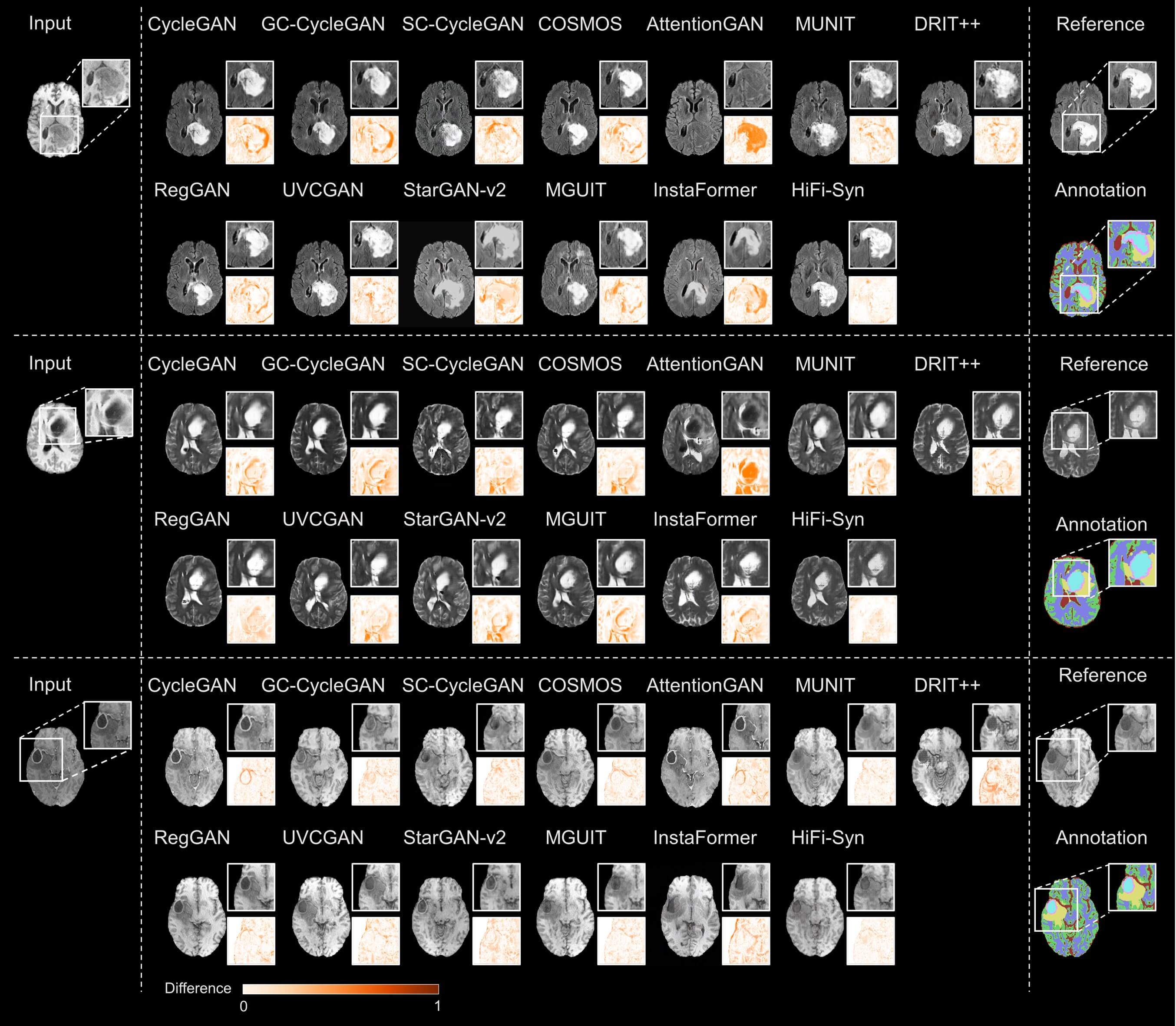}
		\vspace{-0.10in}
	\caption{{Comparison of I2I translation methods on the BraTS dataset.} (From the first to the last sample) T1$\rightarrow$T2-Flair, T1$\rightarrow$T2, and T1ce$\rightarrow$T1 results. Compared to other SOTA methods, HiFi-Syn shows superior structure-preserving translation on both normal and non-normal MR images with clearer error maps. Note that annotations are unseen during testing.}
 
	\label{fig:comparision-tumor}
	\vspace{-0.12in}
\end{figure*}

\begin{table*}[h]   
\centering
\caption{Quantitative evaluation on the BraTS dataset. We perform bidirectional translation for each domain pair and evaluate the performance using PSNR and SSIM metrics, where higher values indicate better results. The p-value is obtained from the Wilcoxon Signed-Rank Test results.}
\vspace{+0.2cm}
\resizebox{1.0\textwidth}{!}{
\begin{tabular}{l|cc cc|cc cc|cc cc|cc|cc}
\hline

\multirow{2}{*}{Method} & \multicolumn{2}{c}{T1 $\rightarrow$ T2} & \multicolumn{2}{c|}{T2 $\rightarrow$ T1} & \multicolumn{2}{c}{T1 $\rightarrow$ T1ce} & \multicolumn{2}{c|}{T1ce $\rightarrow$ T1} & \multicolumn{2}{c}{T1 $\rightarrow$ Flair} & \multicolumn{2}{c|}{Flair $\rightarrow$ T1} & \multicolumn{2}{c|}{Avg} & \multicolumn{2}{c}{P-value} \\ 
                        \cline{2-17}  
                        & PSNR & SSIM  & PSNR & SSIM & PSNR & SSIM & PSNR & SSIM & PSNR & SSIM & PSNR & SSIM & PSNR & SSIM  & PSNR & SSIM \\ \hline \hline
CycleGAN              & 16.4163 & 0.7701 & 18.8426 & 0.7835 & 22.3416 & 0.7952 & 23.9831 & 0.8130 & 16.3494 & 0.7458 & 20.4216 & 0.7878 & 19.7258 & 0.7826 & 7e-20 & 3e-48 \\
GC-CycleGAN           & 16.7513 & 0.7768 & 18.9146 & 0.7894 & 22.3140 & 0.7943 & 23.8754 & 0.8114 & 16.4155 & 0.7512 & 20.4481 & 0.7910 & 19.7865 & 0.7857 & 6e-16 & 6e-35 \\
SC-CycleGAN           & 16.8314 & 0.7801 & 19.1136 & 0.7914 & 22.8474 & 0.8005 & 24.1542 & 0.8155 & 16.7621 & 0.7548 & 20.5253 & 0.7934 & 20.0390 & 0.7893 & 8e-27 & 2e-61 \\
COSMOS                & 18.0308 & 0.7942 & 20.3238 & 0.8049 & 22.9407 & 0.8118 & 24.0575 & 0.8220 & 17.0313 & 0.7671 & 21.3072 & 0.8029 & 20.6152 & 0.8005 & 1e-35 & 3e-43 \\
AttentionGAN          & 19.6376 & 0.8065 & 21.1903 & 0.8151 & 23.1510 & 0.8164 & 24.7227 & 0.8386 & 18.4009 & 0.7742 & 22.2693 & 0.8126 & 21.5620 & 0.8106 & 6e-27 & 2e-13 \\
MUNIT                 & 20.3364 & 0.8167 & 19.5421 & 0.8058 & 22.0965 & 0.7531 & 23.6318 & 0.7736 & \textbf{21.6752} & 0.7670 & 20.6847 & 0.7569 & 21.3278 & 0.7789 & 2e-15 & 7e-28 \\
DRIT++                & 22.3565 & 0.8580 & 24.9558 & 0.8790 & 19.0562 & 0.7914 & 26.0656 & 0.8758 & 18.9193 & 0.7709 & 24.4028 & 0.8614 & 22.6260 & 0.8394 & 9e-15 & 6e-34 \\
RegGAN                & 20.8603 & 0.8283 & 22.1322 & 0.8399 & 21.3708 & 0.8175 & 23.2852 & 0.8447 & 19.3169 & 0.7713 & 21.5751 & 0.8147 & 21.4234 & 0.8194 & 2e-36 & 3e-47 \\
UVCGAN                & 21.8621 & 0.8335 & 22.6261 & 0.8436 & 22.1844 & 0.8242 & 23.3948 & 0.8553 & 19.8276 & 0.7756 & 22.2807 & 0.8215 & 22.0293 & 0.8256 & 1e-28 & 9e-21 \\
StarGAN-v2            & 21.3841 & 0.7704 & 21.9111 & 0.7821 & 21.9245 & 0.8294 & 23.1861 & 0.8464 & 19.1493 & 0.7592 & 21.4930 & 0.7736 & 21.5080 & 0.7935 & 6e-37 & 3e-39 \\
MGUIT                 & 21.9644 & 0.8775 & 25.6549 & 0.8922 & 18.9326 & 0.8180 & 26.1800 & 0.8765 & 19.4815 & 0.7772 & 23.2653 & 0.8231 & 22.5798 & 0.8441 & 9e-51 & 4e-32 \\
InstaFormer           & 22.0187 & 0.8793 & \textbf{25.7539} & 0.8916 & 22.1915 & 0.8204 & 24.9027 & 0.8672 & 20.2528 & 0.7864 & 24.7532 & 0.8671 & 23.3121 & 0.8520 & 8e-26 & 5e-17 \\
HiFi-Syn              & \textbf{23.4402} & \textbf{0.8941} & 25.7228 & \textbf{0.9078} & \textbf{25.6167} & \textbf{0.8551} & \textbf{26.8424} & \textbf{0.8973} & 21.3512 & \textbf{0.8061} & \textbf{26.0135} & \textbf{0.8836} & \textbf{24.8311} & \textbf{0.8740} & - & - \\

\hline
\end{tabular}}
\label{tb:comparision-sota-brats}
\end{table*}

\subsection{Quantitative Evaluation on Tumor-Bearing Samples}
Non-normal structures, specifically referring to tumors in our study, exhibit varying contrasts across different imaging modalities due to their pathological characteristics. These non-normal structures hold significant pathological implications, prompting us to conduct further evaluations. In particular, we assessed the network's capacity to accurately retain the structural characteristics of normal tissues while simultaneously deciphering the anatomical features of non-normal tissues. Therefore, in this subsection, we present evaluations conducted on the BraTS dataset.

The PSNR and SSIM metrics are listed in the Table~\ref{tb:comparision-sota-brats}. We can observe that HiFi-Syn exhibits a substantial lead over other methods, and this advantage becomes even more pronounced when compared to the advantages observed in previous normal brain datasets. This enhanced performance is likely attributed to the intricate semantic information inherent in tumor-bearing images, which may pose difficulties for other methods in discerning various structures. The instance-level I2I network, InstaFormer and MGUIT, achieved results within the second tier. These methods, in contrast to those exclusively concentrating on global content, take into account instance-level representations, which can capture more detailed and local information. However, the challenges lie in the granularity of instance-level learning, which can struggle to effectively handle complex anatomical structures. This limitation becomes evident in scenarios where intricate structures and fine-grained details require a more nuanced comprehension, potentially contributing to their placement in the second tier of results. Through visual inspection of the generated results shown in Fig.~\ref{fig:comparision-tumor}, we find that synthetic images produced by global-level methods like AttentionGAN can create confusion in the appearances of structures across different modalities. For example, MRI images generated by AttentionGAN for some water-rich tissues, such as tumor edema, exhibit bright contrast in T1 and dark contrast in T2, which is the opposite of what is typically observed in real MRI scans.
Besides, other baselines introduce varying degrees of distortion in the tumor regions, potentially compromise downstream tasks.

\begin{figure}[h]
	\centering
	\includegraphics[width=0.40\textwidth]{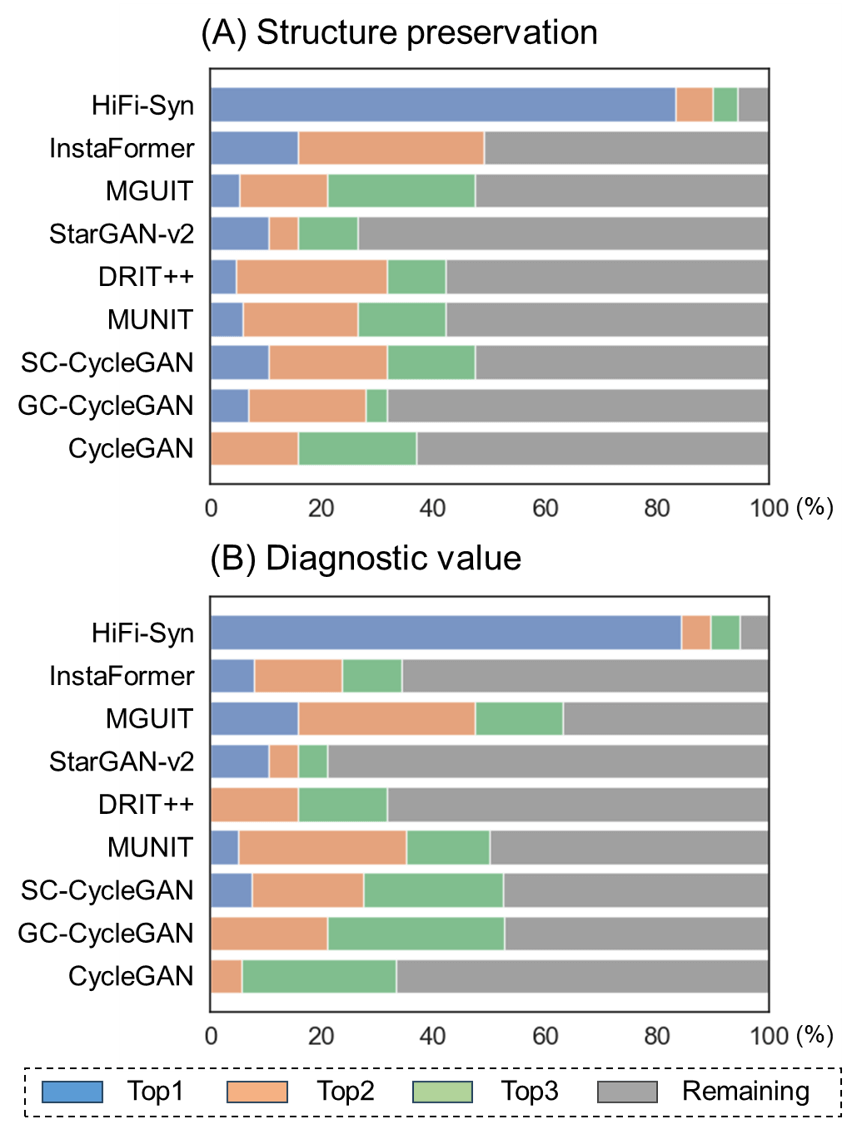}
		\vspace{-0.12in}
	\caption{Expert evaluation results on BraTS dataset. In terms of (a) structural preservation and (b) diagnostic value, HiFi-Syn is scored as the best method for tumor-bearing samples. Experts were allowed to provide tied rankings to different methods.
    } 
	\label{fig:rater-study}
\end{figure}

Furthermore, we conducted an expert study to evaluate the synthesized images generated by different methods. We randomly selected 18 subjects from BraTS test set, encompassing a total of 234 images that underwent evaluation by 6 experts for their structural preservation and diagnostic value. As shown in Fig.~\ref{fig:rater-study}, HiFi-Syn surpasses others with a significant margin for both metrics.

\begin{table}[]
\centering
\caption{Quantitative results for FSL FAST~\citep{jenkinson2012fsl} segmentation using synthetic T1 image.
We report per-class Dice score and volumetric similarity results for the T2$\rightarrow$T1 case via voxel-based analysis. }
\label{tb:fsl-app-metrics}

\resizebox{1.0\linewidth}{!}{
\begin{tabular}{l|c|c|c|c}
\hline
\multirow{2}{*}{Method}&\multicolumn{4}{c}{T2 $\rightarrow$ T1}\\
\cline{2-5}
	&	GM Dice	&	GM VS	&	WM Dice	&	WM VS		\\ \hline\hline
CycleGAN ~\citep{zhu2017cycleGAN}	         &	85.5	&	95.7	&	85.9	&	95.7		\\
GC-CycleGAN ~\citep{R2_hiasa2018cross}	     &	85.7	&	95.9	&	86.1	&	89.2		\\
SC-CycleGAN ~\citep{R2_yang2020unsupervised} &	86.1	&	96.3	&	86.4	&	89.6		\\
COSMOS ~\citep{shin2022cosmos}	             &	86.3	&	96.4	&	86.5	&	89.8		\\
AttentionGAN ~\citep{tang2021attentiongan}	 &	85.2	&	95.2	&	85.4	&	93.5		\\
MUNIT ~\citep{huang2018MUNIT}	             &	86.4	&	96.7	&	87.0	&	96.1		\\
DRIT++ ~\citep{lee2020drit++}	             &	87.3	&	94.1	&	88.3	&	96.6		\\
RegGAN ~\citep{kong2021RegGAN}	             &	86.1	&	96.5	&	86.8	&	95.9		\\
UVCGAN ~\citep{torbunov2023uvcgan}	         &	86.9	&	97.0	&	87.0	&	96.6		\\
StarGAN-v2  ~\citep{choi2020starganV2}       &	86.2	&	95.5	&	84.9	&	96.2	\\
MGUIT ~\citep{jeong2021_MGUIT}	             &	89.2	&	97.2	&	87.5	&	97.0		\\
InstaFormer  ~\citep{kim2022instaformer}     &	89.3	&	96.6	&	88.6	&	97.1	\\
HiFi-Syn                                     &	\textbf{89.9}	&	\textbf{97.5}	&	\textbf{90.3}	& \textbf{98.3}	\\ \hline
\end{tabular}}
\vspace{-5pt}

\end{table}

\begin{figure}[h]
	\centering
	\includegraphics[width=1.0\linewidth]{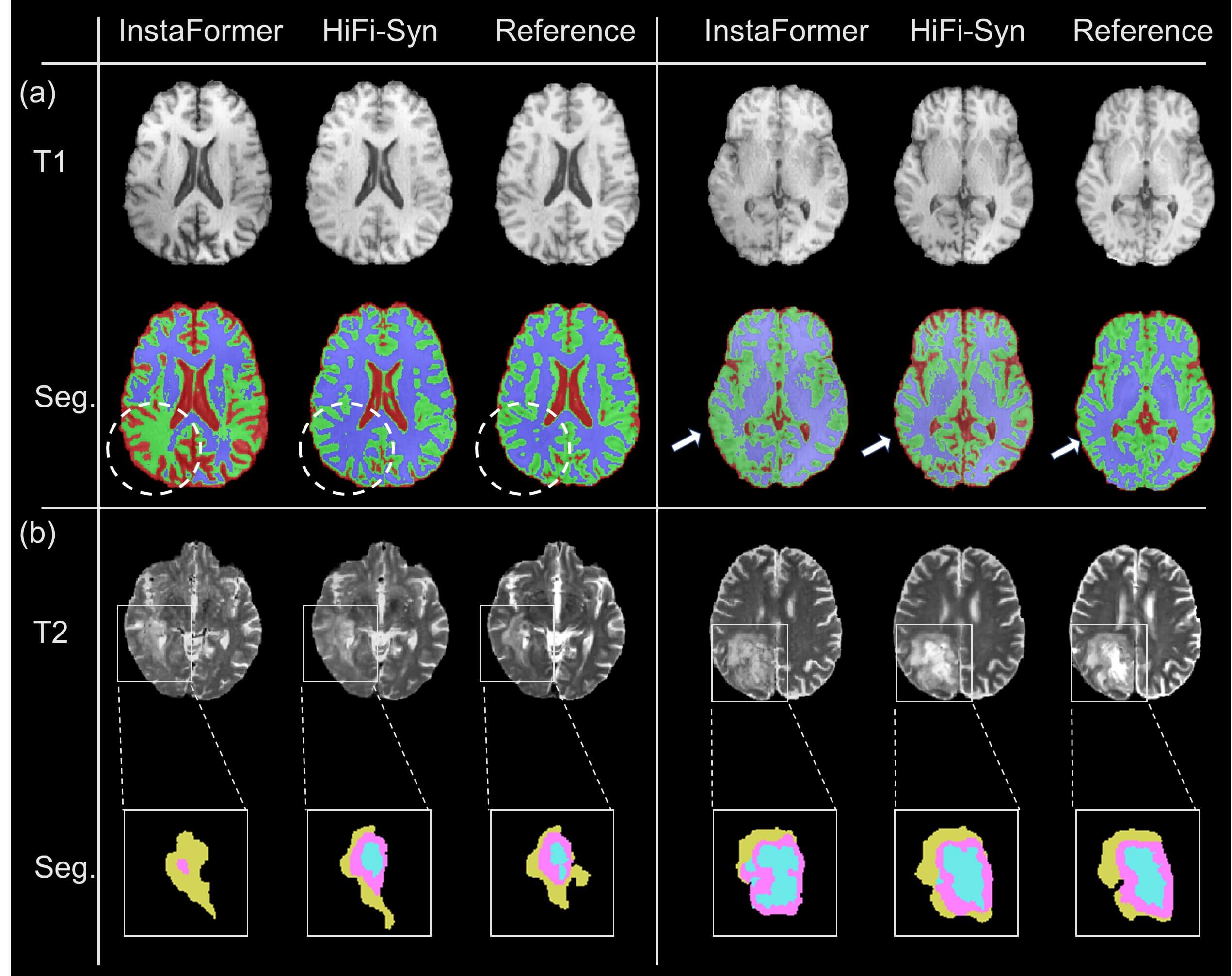}
		\vspace{-0.12in}
	\caption{Synthesis evaluation through downstream (a) FSL FAST and (b) nnUNet segmentation tasks. Compared to InstaFormer (the best SOTA method), the synthetic images from HiFi-Syn retain minimal mismatches in downstream tasks (Green: GM; Blue: WM; Red: CSF; Yellow: Edema ; Pink: Necrosis; Cyan: Enhancing tumor). The arrows and circles in white color indicate some representative regions.} 
	\label{fig:fsl-app-result}
\end{figure}

\begin{figure}[h]
	\centering
	\includegraphics[width=0.48\textwidth]{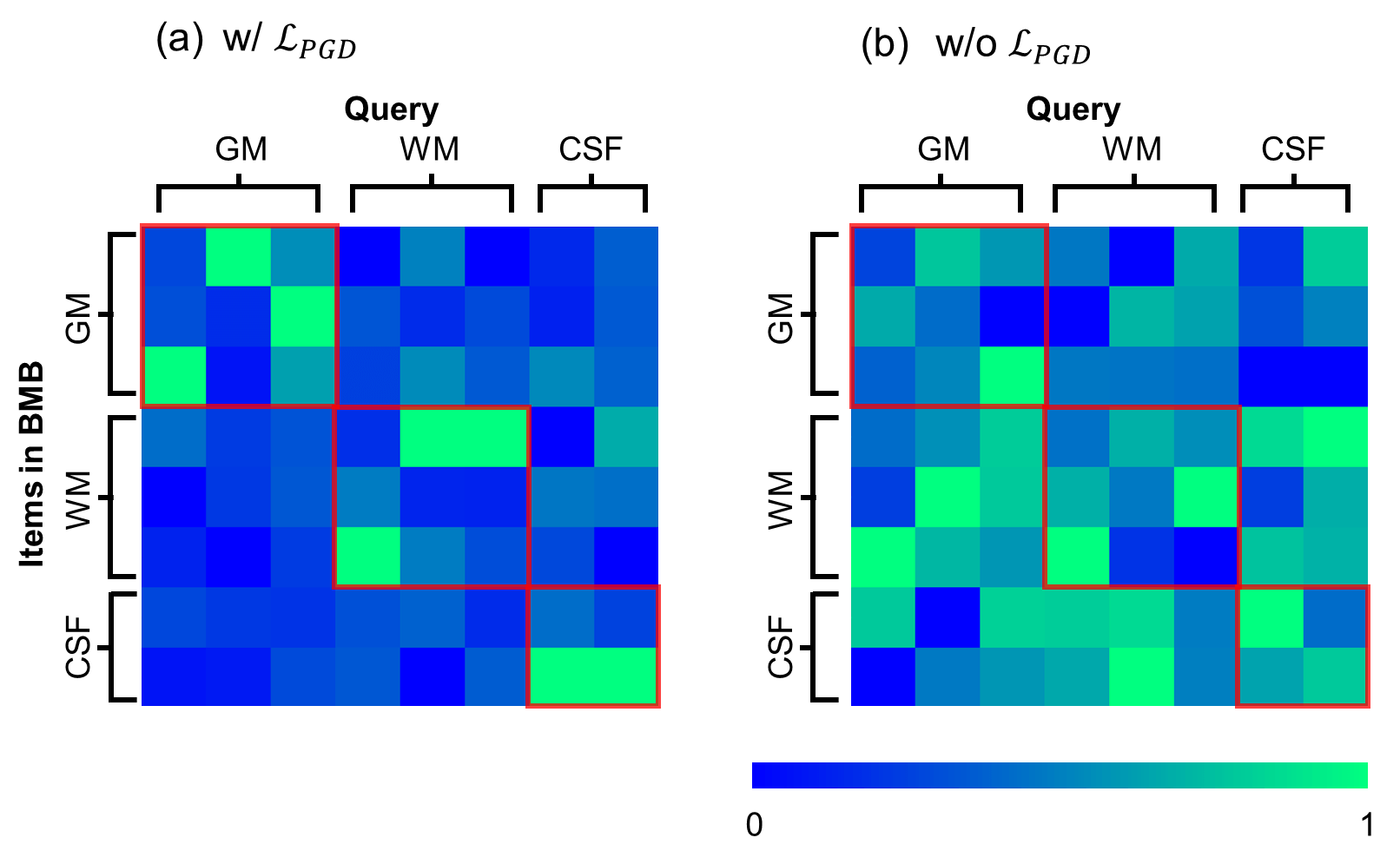}
		\vspace{-0.12in}
	\caption{Visualization of normalized affine scores in Equation ~(\ref{eq:attention-affine-score}). (a) learned w/ and (b) w/o ${\cal L}_{PGD}$. The red box represents reasonable matches between the query and the items (blue: low; green: high). For better visualization, eight randomly selected queries and items from GM, WM, and CSF trained on IXI dataset are involved.
    } 
	\label{fig:Visualization-affine-score}
\end{figure}

\begin{table}[t]
	\centering
	\caption{Improvement comparison on IXI segmentation with different image synthetic methods. }
	\vspace{+0.1in}
 
    \resizebox{0.90\linewidth}{!}{
	\begin{tabular}{l|c|c|c|c}
		\hline
		T1 testing (\%)          & GM & WM & CSF & Avg. \\ \hline \hline

            Real    &   94.3  &   95.1     &   96.2  &  95.2  \\ 

            +Syn. (InstaFormer \citep{kim2022instaformer}) &   94.6  &   95.5     &   96.4  &  95.5  \\ 
		+Syn. (HiFi-Syn)                                   &   \bf 94.9  & \bf   95.8    & \bf  96.5    & \textbf{95.7}  \\ \hline \hline
  
		T2 testing (\%)          & GM & WM & CSF & Avg. \\ \hline

            Real     &   77.6  &   83.9     &   87.8  &  83.1  \\ 
  
		+Syn. (InstaFormer \citep{kim2022instaformer}) &   78.0  &   84.3    &   88.1  &  83.5   \\ 
		  +Syn. (HiFi-Syn)                                   &   \bf 78.5  & \bf   84.6    & \bf  88.3   & \textbf{83.8}  \\ \hline
  
	\end{tabular} }
	\label{tb:IXI-syn-as-augment}
\end{table}

    \setlength{\tabcolsep}{0.15em}
\begin{table}
        \caption{{Quantitative comparison for variants of HiFiSyn.} We measure the average PSNR and SSIM on the IXI dataset. Configuration (\romannumeral1) and (\romannumeral2) are detailed in Section~\ref{sec:ablation-study}. }

\small
    \begin{center}
        \label{tb:Ablation}
        \begin{tabular}{c | c c  c | c c}
            \hline
             \multirow{2}*{\parbox{2em}{\centering Task}}  & \multirow{2}*{\parbox{2.7 em}{\centering $\mathcal{L}_\mathrm{PGD}$}} & \multirow{2}*{\parbox{2.7 em}{\centering $\mathcal{L}_\mathrm{SGD}$}} & \multirow{2}*{\parbox{2.7em}{\centering $\mathcal{L}_\mathrm{GGD}$}}  & \multirow{2}*{\parbox{3.5em}{\centering PSNR }} & \multirow{2}*{\parbox{3.5em}{\centering SSIM }}\\
            &   &  & & & \\
            \hline \hline
            \multirow{6}{*}{$ \text{T1} \rightarrow \text{T2}$} 
             &    -   &  -      &   -   &  22.1485 & 0.7950 \\
            
             &    (\romannumeral1)   &  -      &   -   &  22.1682 & 0.8041 \\
             &    (\romannumeral2)     &   -     & -  &  22.5184 & 0.8063 \\
             & \checkmark       & - & - & 23.4156 & 0.8174 \\
             & \checkmark & \checkmark &    -   &    23.8073   & 0.8212 \\
             & \checkmark & \checkmark & \checkmark & \textbf{23.8376} & \textbf{0.8240} \\
            \hline
            \multirow{6}{*}{ $ \text{T2} \rightarrow \text{T1} $}
             &  -  &    -    & -  & 22.3672  & 0.8093 \\
            
             &  (\romannumeral1)  &    -    & -  & 22.6818  & 0.8164 \\
             &  (\romannumeral2)   &   -     & -  &  22.7679 & 0.8185 \\
             & \checkmark       &  - & - & 23.4603 & 0.8279 \\
             & \checkmark & \checkmark &  -    &   23.6274  & 0.8347 \\
             & \checkmark & \checkmark & \checkmark & \textbf{24.1751} & \textbf{0.8371} \\
            \hline
        \end{tabular}
    \end{center}
\end{table}

\begin{table}[t]
	\centering
	\caption{Comparison of different baseline methods without annotation maps and with annotation maps (via $\mathcal{L}_\mathrm{PGD}$) through pixel-level discrimination on the IXI dataset.}
	\vspace{+0.1in}
 
    \resizebox{1.0\linewidth}{!}
    {
	\begin{tabular}{l|c|c|c|c}
		\hline
		~T1 $\rightarrow$ T2~      & w/o Annot.  & w/ Annot.  & PSNR & SSIM \\ \hline \hline

            MUNIT \citep{huang2018MUNIT} &   \checkmark   &      &   ~20.9800~  &  ~0.7833~  \\ 

            MUNIT \citep{huang2018MUNIT} &     &    \checkmark    &   21.1744 &  0.7956  \\ 
            MGUIT \citep{jeong2021_MGUIT} &  \checkmark    &       &   24.3749  &  0.7949  \\ 
            
		  MGUIT \citep{jeong2021_MGUIT}  &     &   \checkmark    & 24.6809    & 0.8019  \\ \hline \hline
  
		~T2 $\rightarrow$ T1~     & w/o Annot. & w/ Annot. & ~PSNR~ & ~SSIM~ \\ \hline

            MUNIT \citep{huang2018MUNIT} &   \checkmark   &       &   21.7122  &  0.8024  \\ 
  
		MUNIT \citep{huang2018MUNIT} &     &   \checkmark     &   22.4385  &  0.8146   \\ 
		MGUIT \citep{jeong2021_MGUIT} &   \checkmark   &      &   23.2653  &  0.8231   \\ 
  
		  MGUIT \citep{jeong2021_MGUIT} &     & \checkmark      &   23.9366  & 0.8287  \\ \hline
  
	\end{tabular} }
	\label{tb:add-map}
\end{table}

\begin{figure}[h]
	\centering
	\includegraphics[width=0.38\textwidth]{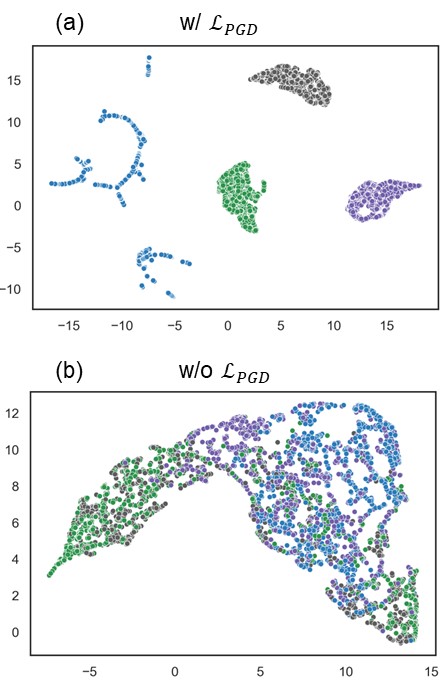}
	\caption{Visualization of attribute space distribution by UMAP \citep{mcinnes2018umap}. (a) w/ and (b) w/o ${\cal L}_{PGD}$  (green: GM; grey: WM; purple: CSF; blue: background).} 
	\label{fig:UMAP-visualization}
\end{figure}

\subsection{Voxel-Level Analysis}
\subsubsection{Following the voxel-based analysis Paradigm}
Structural mismatches, such as those involving gray or white matter, can result in misdiagnosis or the identification of unreliable biomarkers in neuroimaging downstream tasks. Inspired by the voxel-based analysis paradigm~\citep{ashburner2000voxel}, we first segment our synthesized T1 image into gray and white matter in the IXI dataset using FSL FAST~\citep{jenkinson2012fsl}, to assess the amount of the anatomical information transferred by HiFi-Syn from the real image. The results, presented in Fig.~\ref{fig:fsl-app-result}(a) and Table~\ref{tb:fsl-app-metrics}, show that our results are consistently better than other methods as measured by Dice scores and volume similarity (VS) that quantify the overlap of structures between synthetic and real images.

For the T2 synthesized images, we have shifted our focus to comparing tumor region segmentation in the BraTS dataset, as the T2 modality provides rich contrast in tumor regions. We employ nnUNet~\citep{isensee2019nnUNet} for tumor segmentation, utilizing this out-of-the-box solution that has been trained on a subset of data from the corresponding modality, excluding the I2I translation test set. The results presented in Fig.~\ref{fig:fsl-app-result}(b) demonstrate that HiFi-Syn maintains the highest tumor region similarity and more closely approximates the segmentation results obtained with real images compared to the best SOTA method.

\subsubsection{Data Augmentation for Segmentation on IXI Dataset}
We further investigate the impact of image synthesis for data augmentation purpose. For this purpose, we assess the improvement achieved by HiFi-Syn compared to the SOTA generation methods. We have selected the InstaFormer method as the primary baseline due to its best performance among competing methods. We employ nnUNet~\citep{isensee2019nnUNet} to train segmentation models for each individual modality. The GAN models are used to generate synthetic duplicates for the entire T1 and T2 datasets, encompassing 60 subjects in the train set and 20 subjects in the test set. These 60 synthetic images are then integrated into the training stage to serve as a supplement where necessary.

The impact of this integration is detailed in Table~\ref{tb:IXI-syn-as-augment}, with different columns specifying the kinds of the training data employed. It was evident that nnUNet, when trained solely on authentic T1 or T2 images, attained an average Dice score of 95.2\% and 83.1\%, respectively. Notably, the integration of synthetic images from HiFi-Syn into the training set yielded a performance increment, with Dice scores improving by 0.5\% to 0.7\%. This increment surpasses the augmentation strategy observed with InstaFormer, thereby indicating the superior utility of our method. This approach, employing image synthesis as a data augmentation technique, leverages unpaired medical imaging data to enrich the diversity of the training dataset. By incorporating synthetic MRI images, the GAN model could provide an additional information that complements and extends beyond what is available from solely real MRI images. This method, therefore, employs the synthesized data to introduce a higher level of heterogeneity, which in turn improves the robustness and generalizability of the segmentation models.

\subsection{Ablation Study}
In this subsection, we investigate how hierarchical granularity discrimination improves translation performance. We present the following experiment results and analyses from three aspects, the impact of three discrimination strategies, visualization of memory items from BMB as well as latent content feature distribution.

\subsubsection{Impact of Discrimination Strategies}
\label{sec:ablation-study}
We define our framework without the BMB and three discrimination losses as the baseline model and explore the impact of different configurations:
(1) with pixel-level granularity discrimination loss ${\cal L}_{PGD}$ (including both BMB and the annotation map);
(2) with structure-level granularity discrimination loss ${\cal L}_{SGD}$;
(3) with global-level granularity discrimination loss $ {\cal L}_{GGD} $.
For the BMB and annotation map setup, we compared three configurations:
(\romannumeral1) ${\cal L}_{PGD}$ used without the annotation map, which involves using only global memory items in the BMB without structure categories;
(\romannumeral2) with the annotation map during training but without the BMB.
Besides, we also conducted experiments where the annotation map was injected into two baseline models (MUNIT \citep{huang2018MUNIT} and MGUIT \citep{jeong2021_MGUIT}) via ${\cal L}_{PGD}$ to investigate how the additional structural information affect the model performance. This was done to explore how the inclusion of additional structural information impacts model performance.
We execute ablation study on IXI dataset, which is suitable to demonstrate the efficiency of individual units.

The quantitative results are shown in Table~\ref{tb:Ablation}. We can observe that the configuration with ${\cal L}_{PGD}$ demonstrates a notable improvement with the integration of annotation maps into the network, compared to setting (\romannumeral1).
Furthermore, when comparing experiments ${\cal L}_{PGD}$ with and without the BMB, as denoted by (\romannumeral2), it is clear that the memory bank is crucial for enhancing performance, as it smooths the training process by providing a richer contextual backing at the pixel level.

In Table~\ref{tb:add-map}, the results indicate that GAN performance benefits from the inclusion of additional annotation information across various image-to-image translation tasks, regardless of the underlying backbone used.
It is also important to note that the integration of these modules does not significantly increase the time complexity of our implementation in practice.

\subsubsection{BMB Items and Feature Distribution}
To underscore \( {\cal L}_{PGD} \) as one of the key contributions of our paper, we investigate its impact on disentangled learning by presenting affinity scores from Equation (\ref{eq:attention-affine-score}). These scores compare models trained with and without \( {\cal L}_{PGD} \) on the IXI dataset.
As illustrated in Fig.~\ref{fig:Visualization-affine-score}, it is important to note that queries from specific structures $s_i$ show pronounced activation in corresponding memory items, as highlighted within the red-framed diagonal areas. This demonstrates that our proposed method of pixel-level granularity discrimination enhances the discriminative ability of the model within the context of embedding content features, facilitating the extraction of structure-wise knowledge throughout training, guided by the prior of brain anatomical structural maps. The model, when trained with PGD loss, effectively learns features that are more specific to the structure of each query from the embedded content features. 

From the perspective of disentangled representation learning, the representations learned using \( {\cal L}_{PGD} \) are distinct and well-disentangled compared to those from the model where \( {\cal L}_{PGD} \) is not employed. In the latter scenario, the latent representations in memory items tend to contain blended, entangled features, which diminishes their discriminative ability.

We further visualize in Fig.~\ref{fig:UMAP-visualization} the distribution of query features from different structures in IXI dataset. We can observe that with our proposed losses, model can learned more discriminating embedded features.

\subsection{Potential Clinical Impact}
In contrast to most existing methods that neglect the spatial coherence of structural-level representation and the anatomical consistency of content during model training, our approach addresses the structure-preserving challenge by exploring a novel direction that utilizes hierarchical granularity discrimination to generate high-fidelity images. 
To fully understand the potential clinical impact of our method, we present the following discussions from two aspects: the benefits of imputing missing modalities and data augmentation strategies for imbalanced and skewed clinical datasets.

\subsubsection{Imputing Missing Modalities}

Imputing missing modalities in medical imaging datasets tackles several critical challenges in clinical settings. These challenges include patients unable to fully cooperate, emergencies that require rapid decision-making, and the use of historical data that may be incomplete.

\noindent{\textbf{Challenges in Data Collection.}}
Collecting multi-modal data can be challenging when patients are unable to fully cooperate with imaging protocols due to physical discomfort, anxiety, or other conditions. Additionally, completing all necessary imaging examinations during the initial visit may not always be feasible, and scheduling subsequent sessions to acquire missing data can prove challenging or impractical due to difficulties in contacting patients. Imputing missing modalities under structure-preserving manner with HiFi-Syn can allow clinicians to obtain a comprehensive view of the patient's condition without requiring additional scans. This not only facilitates ongoing care but also ensures that datasets are complete for future analysis and may even offer available solutions for retrospective studies that were previously limited by incomplete data.

\noindent{\textbf{Emergencies and Time-Sensitive Situations.}}
During emergencies or urgent surgical interventions, there often is not enough time to perform lengthy imaging procedures without risking patient safety or delaying critical care. In such situations, the ability to quickly generate synthetic equivalents of missing imaging modalities could be lifesaving. By imputing these missing images, experts can make informed decisions faster, enhancing the chances of successful outcomes.

\noindent{\textbf{Enhancement of Historical Databases.}}
Historical medical databases often contain incomplete records due to the unavailability of certain imaging technologies at the time of data collection or due to partial data collection practices. Imputing missing modalities in such databases enriches these resources, making them more valuable for research and training purposes. Enhanced datasets can help in retrospective studies and improve the training of machine learning models by providing a more comprehensive dataset.

\subsubsection{Data Augmentation Strategies}
Besides, HiFi-Syn can serve as a powerful form of data augmentation, especially useful in the context of imbalanced or skewed clinical datasets. By generating high-quality, multi-modality examples of disease cases from diverse source data, this model can enrich datasets. This enhancement supports the development of more generalized and robust diagnostic algorithms, which is particularly beneficial in machine learning applications where model performance often suffers due to the uneven distribution of data from clinical centers. Expanding datasets with synthetically generated, yet realistic examples helps to mitigate these issues by providing a more balanced representation of rare and common conditions alike, thus improving the accuracy and reliability of predictive models across varied clinical scenarios.

\subsubsection{Limitation and Future Work}  
\noindent{\textbf{Limitation.}}  
While HiFi-Syn demonstrates remarkable capabilities in synthesizing target domain images, it is inescapable to acknowledge its limitations. Like most of recently translation models, such as Syndiff~\citep{ozbey2023syndiff} and ResViT~\citep{dalmaz2022resvit}, the primary constraint resides in its inherent 2D character. Fortunately, the proposed BMB in our framework could alleviate this issue to a certain extent, as it can learn and record the prototypical information across MR slices. Our model continues to demonstrate commendable performance in 3D applications, as illustrated in Fig.~\ref{fig:fsl-app-result}. Besides, recent studies have shown that the 3D versions of 2D GAN models do not significantly surpass the performance of their 2D counterparts in some tasks \citep{hadzic2024optimizing, salmanpour2024high}. To address this limitation, future research directions could explore the integration of 3D spatial information into existing synthesis framework effectively. By doing so, we can further bolster the model's generalization capabilities and its overall value in clinical utility and diagnostic processes.

\noindent{\textbf{Future Work.}}
In light of the findings mentioned above, we have also finded several promising directions for future research.
(i) Evaluation of clinical impact. Further studies are needed to evaluate the clinical impact of generated images, particularly in terms of diagnostic accuracy and patient outcomes. Collaborations with clinical experts could provide valuable insights into the utility and limitations of GAN-based model in real-world medical practice. (ii) Robustness against cross-center data heterogeneity. Assessing the model’s robustness against cross-center data heterogeneity, including variations due to different imaging equipment and protocols, would be a significant step toward ensuring its applicability across diverse clinical environments.

\section{Conclusion}
In this work, we present a hierarchical granularity discrimination strategy designed to leverage the diverse levels of semantic information present in medical images. Our model, HiFi-Syn, employs three levels of discrimination granularity, thereby ensuring anatomical consistency throughout the translation process. The experimental results demonstrate that HiFi-Syn achieves a SOTA performance across three brain MRI datasets, whether in normal or abnormal MR images. The diagnostic value of synthesized MR images containing brain tumors has been assessed by radiologists. This indicates that our model may offer an alternative solution in scenarios where specific MR modalities are unavailable. HiFi-Syn improves image translation capabilities in the context of medical imaging while preserving structural integrity during translation, providing novel insights into the field.

\section{Acknowledgement}
This work was supported in part by grants from the National Natural Science Foundation of China (82171903) and the National Key Research and Development Program of China (2023YFC2410903).

\bibliographystyle{model2-names.bst}\biboptions{authoryear}
\bibliography{medima-template}




\end{document}